\documentclass[]{nature}
\makeatletter\if@twocolumn\PassOptionsToPackage{switch}{lineno}\else\fi\makeatother

\usepackage{tabulary,graphicx,amsmath,amsfonts,amssymb}
\usepackage[utf8x]{inputenc}
\usepackage[T1]{fontenc}

\makeatletter
\renewenvironment{figure}
               {\@float{figure}}
               {\end@float}
\renewenvironment{figure*}
               {\@dblfloat{figure}}
               {\end@dblfloat}

\renewenvironment{table*}
               {\@dblfloat{table}}
               {\end@dblfloat}

\makeatother

\usepackage{url,multirow,morefloats,floatflt,cancel,tfrupee}
\makeatletter

\AtBeginDocument{\@ifpackageloaded{textcomp}{}{\usepackage{textcomp}}}
\makeatother
\usepackage{colortbl}
\usepackage{xcolor}
\usepackage{pifont}
\usepackage[nointegrals]{wasysym}
\makeatletter

\def\mcWidth#1{\csname TY@F#1\endcsname+\tabcolsep}

\def\cAlignHack{\rightskip\@flushglue\leftskip\@flushglue\parindent\z@\parfillskip\z@skip}
\def\rAlignHack{\rightskip\z@skip\leftskip\@flushglue \parindent\z@\parfillskip\z@skip}

\@ifundefined{etal}{}{}

\usepackage{ifxetex}
\ifxetex\else\if@twocolumn\@ifpackageloaded{stfloats}{}{\usepackage{dblfloatfix}}\fi\fi

\AtBeginDocument{
\expandafter\ifx\csname eqalign\endcsname\relax
\def\eqalign#1{\null\vcenter{\def\\{\cr}\openup\jot\m@th
  \ialign{\strut$\displaystyle{##}$\hfil&$\displaystyle{{}##}$\hfil
      \crcr#1\crcr}}\,}
\fi
}

\AtBeginDocument{%
  \@ifpackageloaded{endfloat}%
   {\renewcommand\efloat@iwrite[1]{\immediate\expandafter\protected@write\csname efloat@post#1\endcsname{}}}{\newif\ifefloat@tables}%
}%

\def\BreakURLText#1{\@tfor\brk@tempa:=#1\do{\brk@tempa\hskip0pt}}
\let\lt=<
\let\gt=>
\def\processVert{\ifmmode|\else\textbar\fi}

\@ifundefined{subparagraph}{
\def\subparagraph{\@startsection{paragraph}{5}{2\parindent}{0ex plus 0.1ex minus 0.1ex}%
{0ex}{\normalfont\small\itshape}}%
}{}

\newcommand\role[1]{\unskip}
\newcommand\aucollab[1]{\unskip}
  
\@ifundefined{tsGraphicsScaleX}{\gdef\tsGraphicsScaleX{1}}{}
\@ifundefined{tsGraphicsScaleY}{\gdef\tsGraphicsScaleY{.9}}{}
\def\checkGraphicsWidth{\ifdim\Gin@nat@width>\linewidth
	\tsGraphicsScaleX\linewidth\else\Gin@nat@width\fi}

\def\checkGraphicsHeight{\ifdim\Gin@nat@height>.9\textheight
	\tsGraphicsScaleY\textheight\else\Gin@nat@height\fi}

\def\fixFloatSize#1{}
\let\ts@includegraphics\includegraphics

\def\inlinegraphic[#1]#2{{\edef\@tempa{#1}\edef\baseline@shift{\ifx\@tempa\@empty0\else#1\fi}\edef\tempZ{\the\numexpr(\numexpr(\baseline@shift*\f@size/100))}\protect\raisebox{\tempZ pt}{\ts@includegraphics{#2}}}}

\AtBeginDocument{\def\includegraphics{\@ifnextchar[{\ts@includegraphics}{\ts@includegraphics[width=\checkGraphicsWidth,height=\checkGraphicsHeight,keepaspectratio]}}}

\DeclareMathAlphabet{\mathpzc}{OT1}{pzc}{m}{it}

\def\URL#1#2{\@ifundefined{href}{#2}{\href{#1}{#2}}}

\def\UrlOrds{\do\*\do\-\do\~\do\'\do\"\do\-}%
\g@addto@macro{\UrlBreaks}{\UrlOrds}

\edef\fntEncoding{\f@encoding}

\makeatother

\newif\ifmultipleabstract\multipleabstractfalse%
\usepackage[nomarkers,tablesfirst,nolists]{endfloat}

\def\fixFloatSize#1{}
\usepackage{float}

\newcommand{\texttildeapprox}{{\fontfamily{pcr}\selectfont\texttildelow}}

\makeatletter
\def\mathalfa@calscaled{s*[1] }

  \DeclareFontFamily{U}{BOONDOX-cal}{\skewchar \font =45}
  \DeclareFontShape{U}{BOONDOX-cal}{m}{n}{
    <-> \mathalfa@calscaled  BOONDOX-r-cal}{}
  \DeclareFontShape{U}{BOONDOX-cal}{b}{n}{
    <-> \mathalfa@calscaled  BOONDOX-b-cal}{}
  \DeclareMathAlphabet{\smallmathcal}{U}{BOONDOX-cal}{m}{n}
  \SetMathAlphabet{\smallmathcal}{bold}{U}{BOONDOX-cal}{b}{n}
  \DeclareMathAlphabet{\smallmathbcal} {U}{BOONDOX-cal}{b}{n}
\makeatother

\begin{document}

\title{Bayesian-Deep-Learning Estimation of Earthquake Location from Single-Station Observations}

\author{S.Mostafa Mousavi$^{1}$\thanks{E-mail: mmousavi@stanford.edu}~ \&
              Gregory C. Beroza$^{1}$\thanks{E-mail: beroza@stanford.edu}~
    }

\maketitle 

\begin{affiliations}
  \item 
    Geophysics Department\unskip, 
    Stanford University \unskip, Stanford\unskip, California\unskip, USA
\end{affiliations}

\begin{abstract}
 We present a deep learning method for single-station earthquake location, which we approach as a regression problem using two separate Bayesian neural networks. We use a multi-task temporal-convolutional neural network to learn epicentral distance and P travel time from 1-minute seismograms. The network estimates epicentral distance and P travel time with absolute mean errors of 0.23 km and 0.03 s respectively, along with their epistemic and aleatory uncertainties. We design a separate multi-input network using standard convolutional layers to estimate the back-azimuth angle, and its epistemic uncertainty. This network estimates the direction from which seismic waves arrive to the station with a mean error of 1 degree. Using this information, we estimate the epicenter, origin time, and depth along with their confidence intervals. We use a global dataset of earthquake signals recorded within 1 degree ({\texttildeapprox}112 km) from the event to build the model and to demonstrate its performance. Our model can predict epicenter, origin time, and depth with mean errors of 7.3 km, 0.4 second, and 6.7 km respectively, at different locations around the world. Our approach can be used for fast earthquake source characterization with a limited number of observations, and also for estimating location of earthquakes that are sparsely recorded - either because they are small or because stations are widely separated.

\end{abstract}\def\keywordstitle{Keywords}

    \textbf{\keywordstitle:} AI, Bayesian Neural Networks, Earthquakes, Source Characterization

\section{Introduction}
Recent years have seen a renewed surge of interest in applying machine-learning techniques to seismic signal processing and earthquake monitoring. Promising results from multiple studies indicate neural-network-based models can outperform traditional algorithms in tasks such as: earthquake signal detection \unskip~\cite{542319:14121517,542319:14121518}, phase picking \unskip~\cite{542319:14121519,542319:14121520,542319:14121521,542319:14121522}, first-motion polarity determination \unskip~\cite{542319:14121524,542319:14121525}, denoising \unskip~\cite{542319:14121526,542319:14121527}, discrimination\unskip~\cite{542319:14121528,542319:14121529}, and association\unskip~\cite{542319:14121530,542319:14121531,542319:14121532}. 

However, earthquake location remains a challenging task. Perol et.al.\unskip~\cite{542319:14121517} trained a convolutional neural network to simultaneously learn classification (event vs. noise) and to group earthquakes into 6 clusters initially defined by K-means in Oklahoma. Lomax et.al. \unskip~\cite{542319:14121535} expanded this approach by classifying seismic waveforms into a larger number of classes including: event/noise (1 class), station{\textendash}event distance (50 classes), station{\textendash}event azimuth (36 classes, each 10 degrees), event magnitude (20 classes), and event depth (20 classes); however, their model did not generalize well and suffered from high error rates. On the other hand multi-station approaches (e.g. \unskip~\cite{542319:14403038}) result in a better performance by learning the move out patterns for specific station configuration at a local region. 

Neither these, nor other, neural networks applied to earthquake data quantify uncertainty in their output. Machine learning can be thought of as inferring plausible models that explain data and can be used to make predictions about unseen data.  Uncertainty plays a key role in that process of quantifying the reliability of those predictions. Data can be consistent with different models and the question of what model is appropriate based on such data is uncertain. Predictions using future data are also uncertain \unskip~\cite{542319:12461547}. Typical deep learning models do not capture uncertainties in the output. In regression models, output is a single vector that regresses to the mean of the data, but in classification models, and the output probability is not equivalent to model confidence. That is, a model can be uncertain in its predictions even with a high softmax output\unskip~\cite{542319:12459951}. With model confidence we can treat uncertain input and special cases properly. In the case of classification models for earthquake signal detection and arrival time measurement, for example, we might pass uncertain cases to a human analyst for expert analysis. 

Neural networks with model uncertainty are known as Bayesian neural networks. They offer a probabilistic interpretation of models usually by placing prior probability distributions over the network weights. 

In this paper, we approach single-station earthquake location as a regression problem using two separate Bayesian neural networks. For learning epicentral distance and P travel time we designed a multi-task temporal convolutional network. The network consists of causal dilated convolutions and residual connections that estimate epicentral distance and P travel time simultaneously along with their epistemic and aleatory uncertainties. We use a separate multi-input network with standard convolutional layers to estimate the back-azimuth and its epistemic uncertainty. Using this information, we estimate the epicenter, origin time, and depth along with their confidence intervals. We use a global data set for building the model and for demonstrating its performance. The proposed approach can be used for rapid earthquake source characterization using a limited number of observations. This can have many different applications including in earthquake early warning systems\unskip~\cite{542319:14121746} and in locating earthquakes that are sparsely recorded.

\section{Results}
\textbf{ Dataset}

We use the STanford EArthquake Dataset (STEAD)\unskip~\cite{542319:14404950} for the training and testing of the models. STEAD is a global dataset of labeled 3-component seismic waveforms (earthquake and non-earthquake). Here, we only use earthquake waveforms recorded at epicentral distances of less than 110 km with signal-to-noise ratio of 25 decibels and higher. We only use stations for which north-south and east-west components are properly aligned to their correct geographic orientations. Based on these criteria, we select {\texttildeapprox} 150,000 waveforms to be used for the training (\% 80) and testing (\% 20) of the networks. We show the geographical distribution of the events associated with these waveforms and their characteristics Figure~\ref{f-8b380f12d512} and Figure~\ref{f-75d975c7f33a}. Waveforms are 1 minute in duration with a sampling rate of 100 HZ and were band-passed filtered from 1-45 HZ.

\bgroup
\fixFloatSize{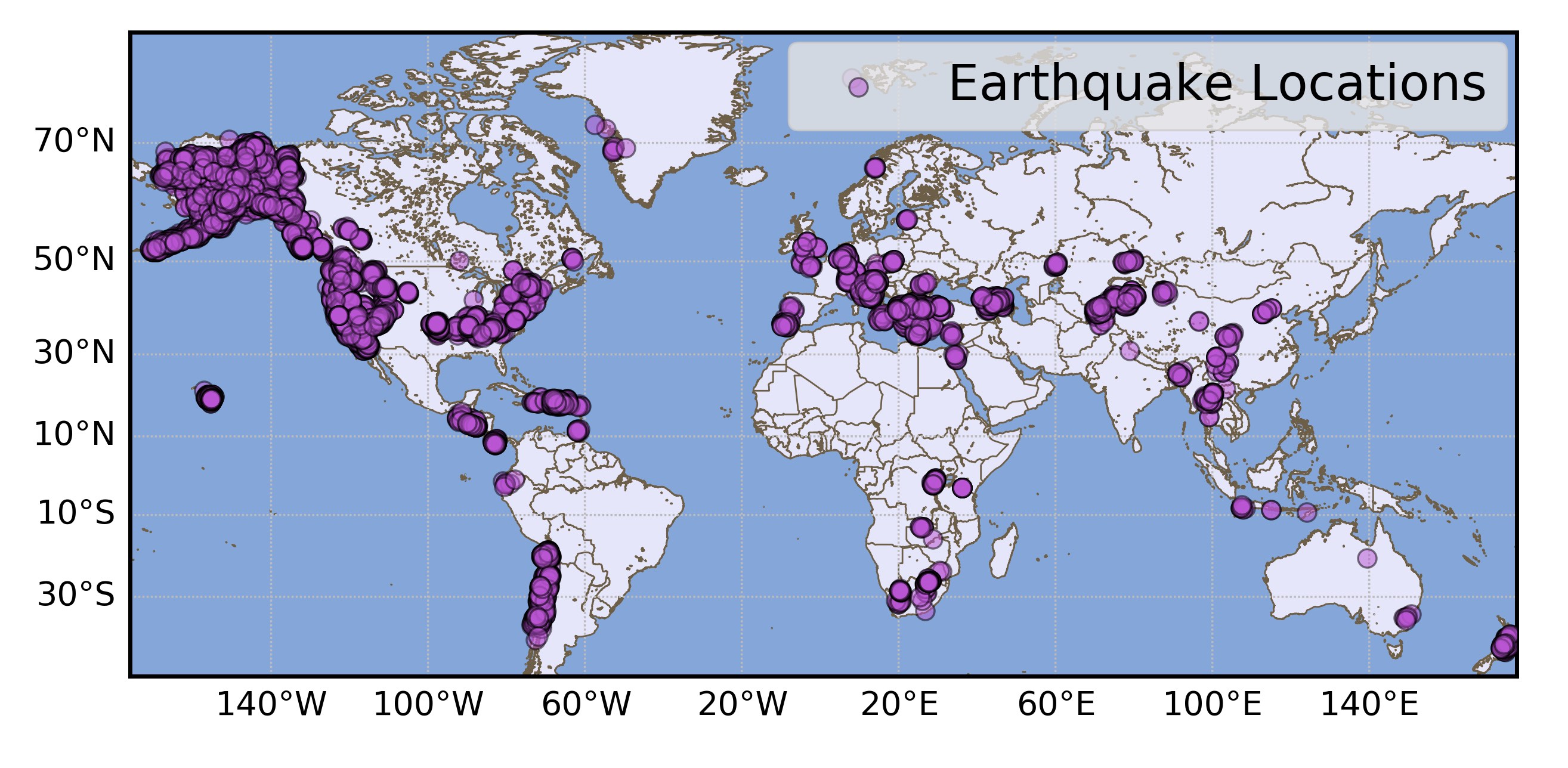}
\begin{figure*}[!htbp]
\centering \makeatletter\IfFileExists{images/567803bd-62f1-48a3-9be1-063acf5ad8e8-ufig_1.jpg}{\includegraphics{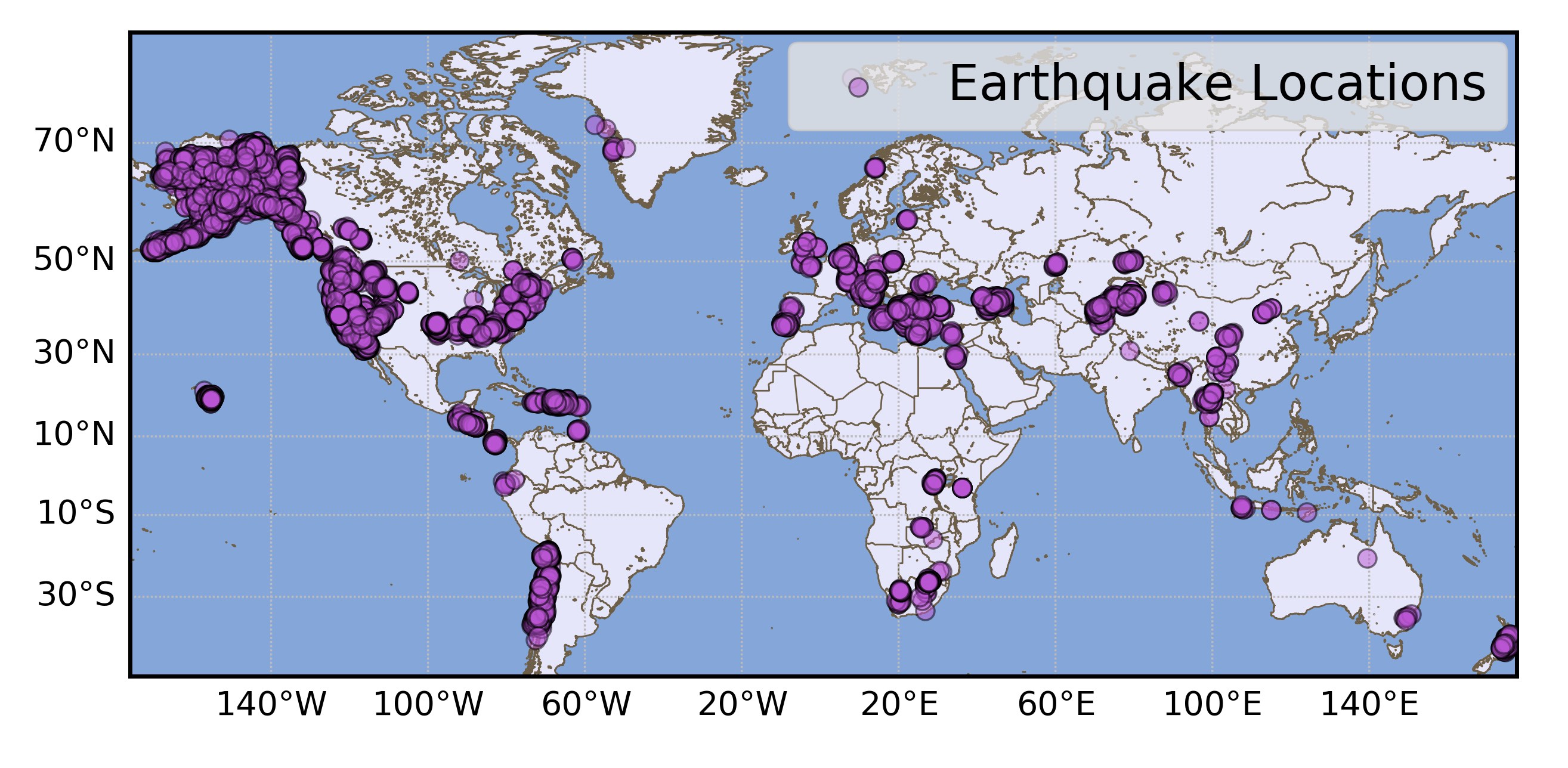}}{}
\makeatother 
\caption{{Geograpical distribution of events used in this study. }}
\label{f-8b380f12d512}
\end{figure*}
\egroup

\bgroup
\fixFloatSize{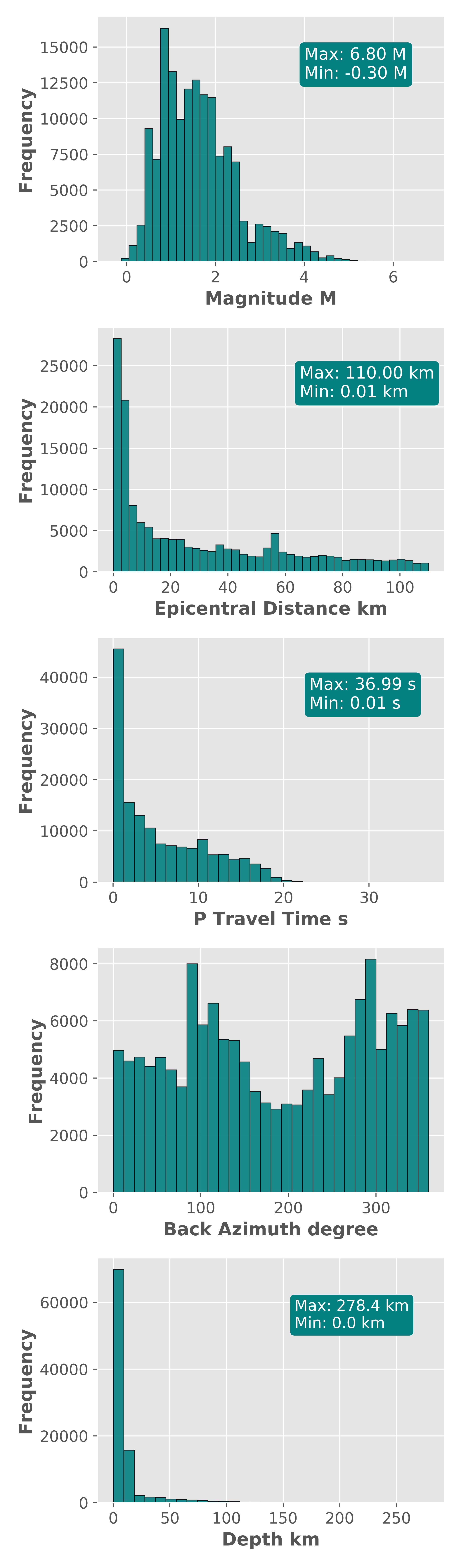}
\begin{figure}[!htbp]
\centering \makeatletter\IfFileExists{images/446cc021-bf33-4043-884f-6b10ef269deb-ufig_2.jpg}{\includegraphics{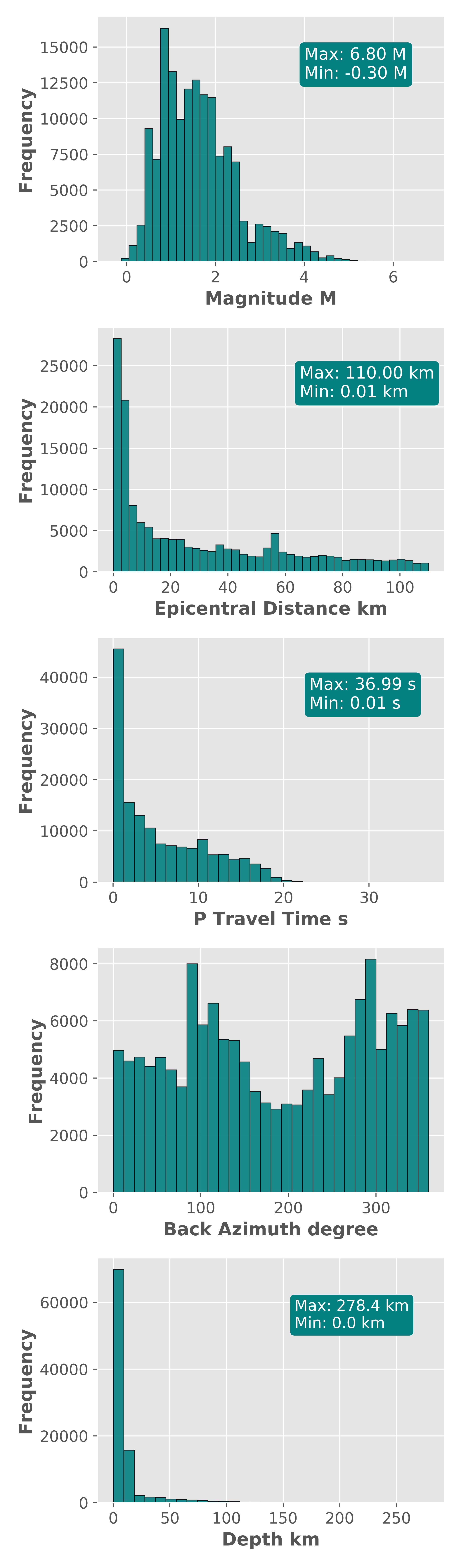}}{}
\makeatother 
\caption{{Characteristics of the dataset used for training and testing. }}
\label{f-75d975c7f33a}
\end{figure}
\egroup
\textbf{ Network Architecture}

We designed two separate networks, one for predicting the epicentral distance and P travel time (dist-PT network) and the other for back-azimuth estimation (BAZ network). 

The dist-PT network is a multi-task temporal convolutional network consisting of 1D convolutional layers where convolutions are causal and dilated Figure~\ref{f-3b0704fbde3f}. The input to the network is a $6000\;\times\;4 $ matrix where the first three rows are 3-component waveforms (each 1 minute long with 100 samples per second) and the last row is a binary vector where values between P and S arrival times are set to 1 and the rest to zero. The last vector is similar to the labeling in output layers of detector/picker networks; however, here we use it as the input to highlight the part of the waveforms that contain the most important information for the regression tasks. 

The main body of the dist-PT network consists of 11 dilational convolution layers (dilation rate doubles for each layer) each with a relu activation function and 20 kernels of size 6. At the end, network has two fully connected layers each with a linear activation function and two neurons. The network has 58,500 trainable parameters. The full description of the optimization function is given in the method section. We applied dropout to every dilated convolutional layer in the network and trained the model with a dropout rate of 0.20. The aleatory uncertainties are implicitly learned from a customized loss function during the training without a need for uncertainty labels. This loss function acts as an intelligent regression function that makes the model robust to noisy data. We sample the posterior distribution over the weights to obtain the posterior probabilities and to estimate uncertainties.

\bgroup
\fixFloatSize{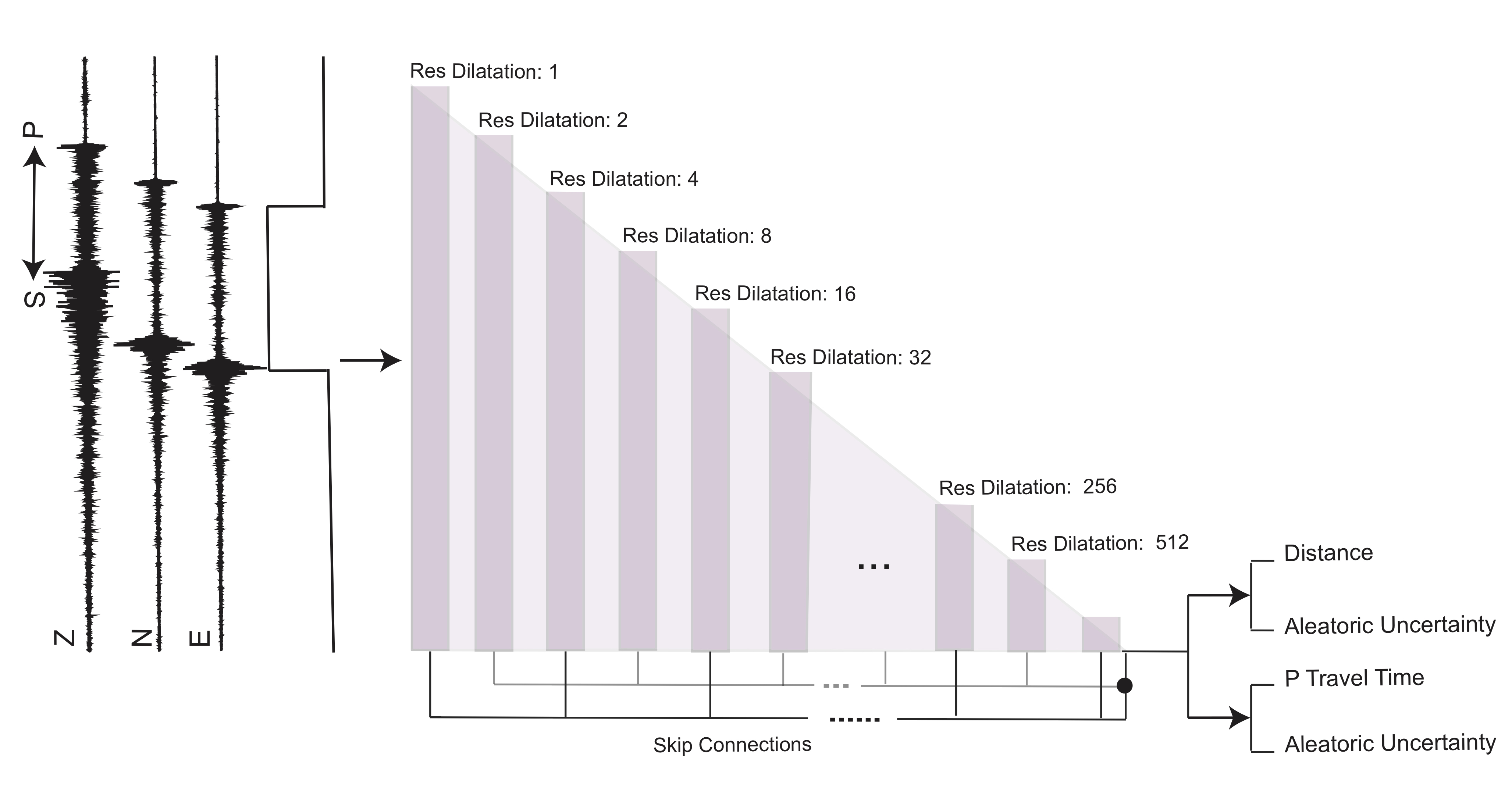}
\begin{figure*}[!htbp]
\centering \makeatletter\IfFileExists{images/beb91c5a-c595-4b29-8b34-0d60eaef34ed-ufig_3.jpg}{\includegraphics{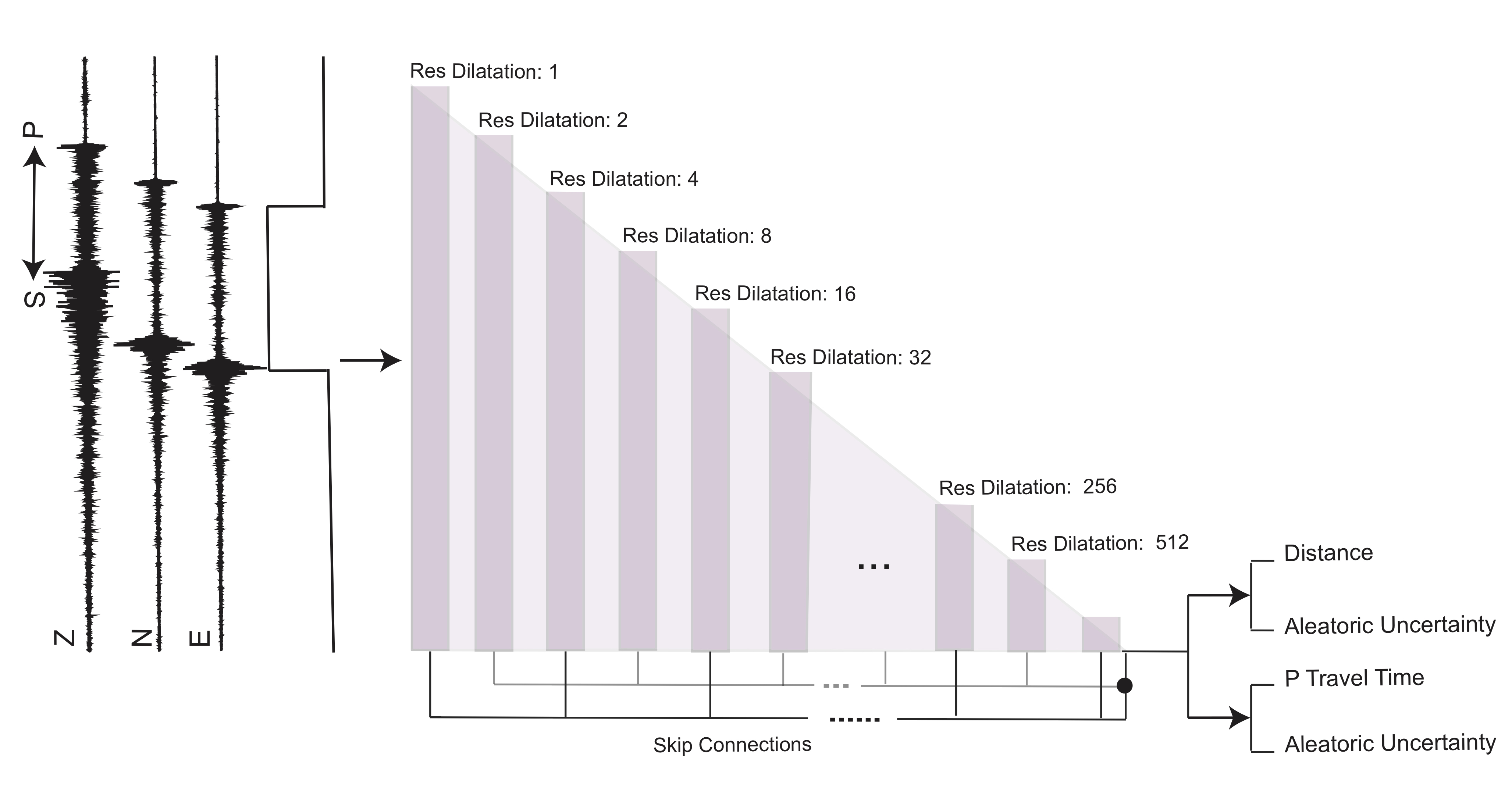}}{}
\makeatother 
\caption{{The dist-PT network for estimating the epicentral distance, P travel time, and their aleatory uncertainties. A detailed description of residual dilational units are presented in the method section and figure Figure~\ref{f-13007899a493}.}}
\label{f-3b0704fbde3f}
\end{figure*}
\egroup
Back-azimuth estimation is a continuous orientation prediction, which prohibits the direct use of a typical L2 loss function because the angle is in a non-Euclidean space. To handle this problem, we represent back-azimuth angles, \textit{baz},  as points on a unit circle $baz\;=\;(\cos\;\theta,\;\sin\;\theta) $ during the training, and convert the predicted 2D points to the back-azimuth angles during testing. 

The back-azimuth network (Figure~\ref{f-6bb8af5bdfa1}) primarily consists of 1D convolutional layers and has two inputs: 1) a $(150\;\times\;3) $ matrix (0.5 second before and 1 second after the P arrival); 2) a $(7\times3) $ matrix consisting of the covariance matrix, eigenvalues, and the eigenvectors derived from the 3-component waveforms for the same time window. We convolve the two input matrices with 4 and 1 convolutional layers and feed them into two fully connected layers with 100 and 2 neurons respectively to predict the coordinates of the back-azimuth angle on the unit circle. All the other layers have relu activations, except for the last fully connected layer. The kernel size used in all convolutional layers is 3 while the number of kernels varies between 20 and 64. Overall, the network is very light and only has {\texttildeapprox}46,000 trainable parameters. Using multiple inputs and point estimations prevent us from a stable estimation of aleatory uncertainty using the intelligent loss function  (Equation~(\ref{dfg-426f090e9702})). This loss function estimates uncertainties for each output ($\cos\;\theta $and $\sin\;\theta $) separately. Moreover, it is hard to assign the estimated uncertainties to the corresponding input. Our attempts to estimate a single aleatory uncertainty for both components did not in a stable optimization. However, we estimate the epistemic uncertainty using the Monte Carlo dropout sampling procedure described in the method section and as done for the previous network. We use dropout rates of 0.1 and 0.3 after the convolutional and fully connected layers respectively.

\bgroup
\fixFloatSize{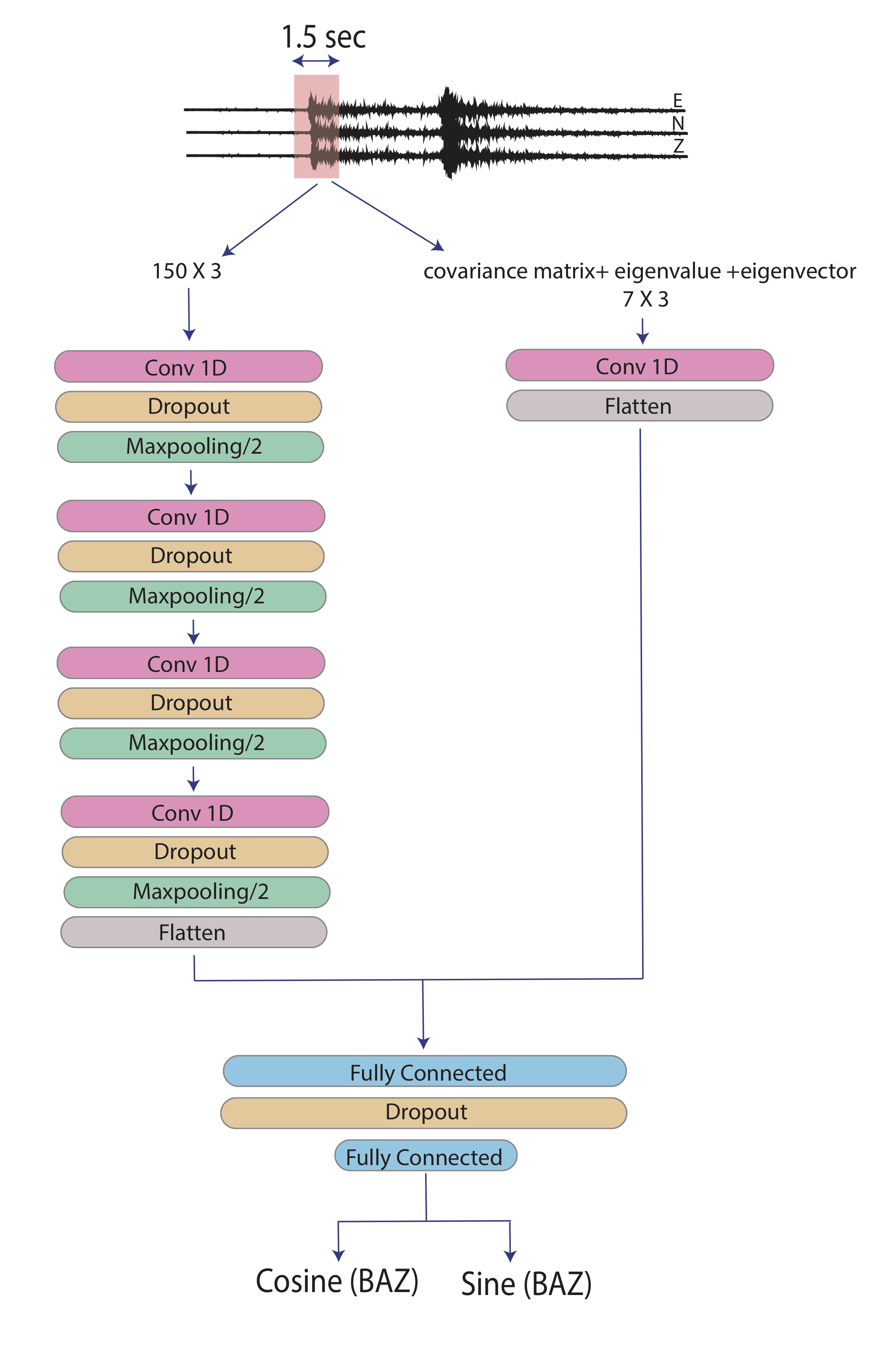}
\begin{figure*}[!htbp]
\centering \makeatletter\IfFileExists{images/7f4839f8-61b4-42ac-aa1d-20204d7b6430-ufig_4.jpg}{\includegraphics{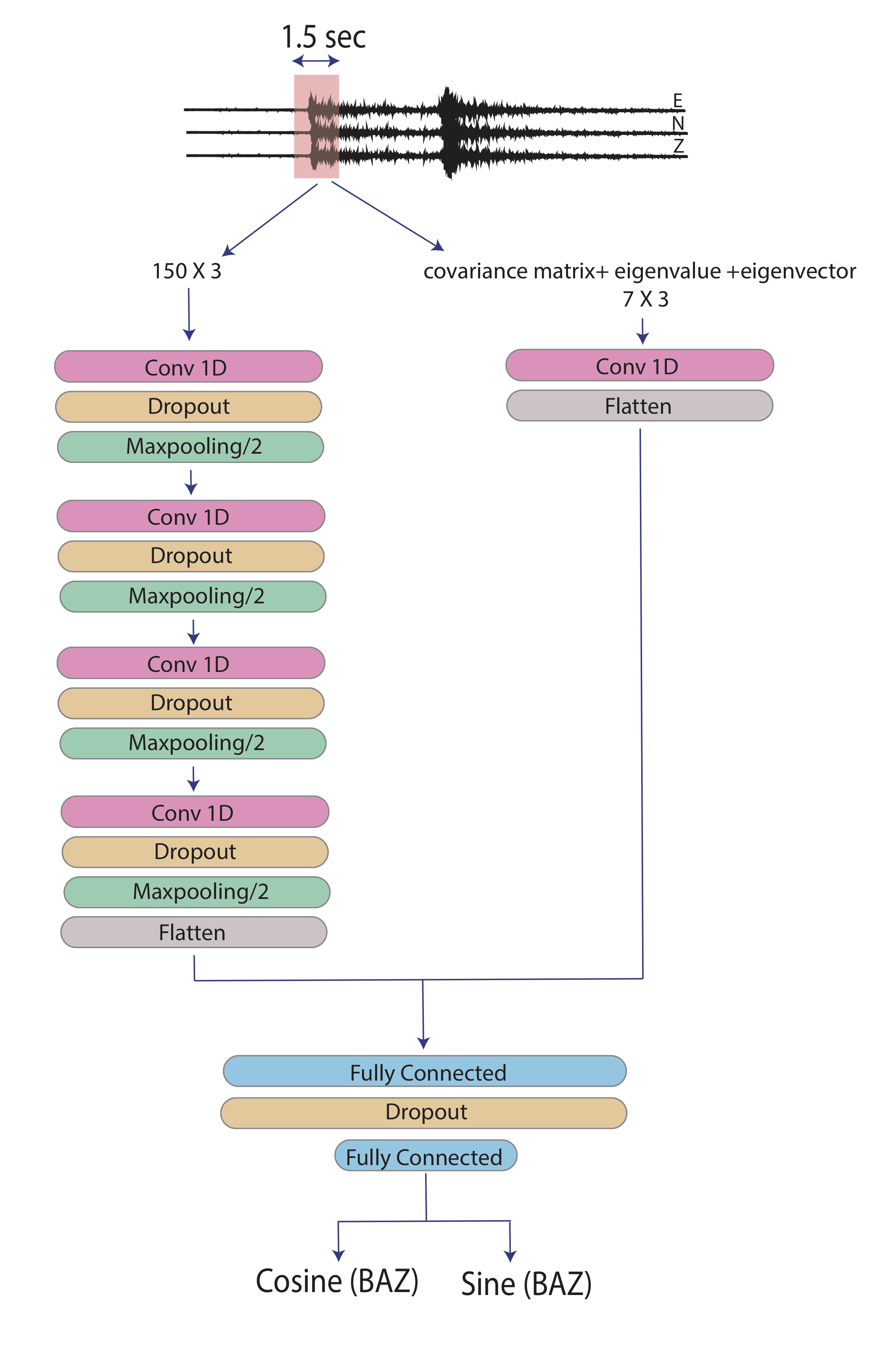}}{}
\makeatother 
\caption{{Architecture of the BAZ network (for estimation of the back azimuth angle (BAZ)). }}
\label{f-6bb8af5bdfa1}
\end{figure*}
\egroup
\textbf{Regression Results}

The regression results for the test set are presented in Figure~\ref{f-5fab5bbf4c0e}. The best coefficient of determination is obtained for P travel time estimation. The network is able to estimate P travel time with a standard deviation of 0.66 second. The mean error for epicentral distance estimates is 0.23 km with a standard deviation of 5.42 km. Compared to these results, back-azimuths estimates are more uncertain. This is mainly due to the complication of estimating orientation; however, a coefficient of determination of 0.87 for the regression results and a mean error rate of {\texttildeapprox} 1 degree can be considered good results due to the fact that only 1.5 seconds of the waveforms are used for these estimates.

\bgroup
\fixFloatSize{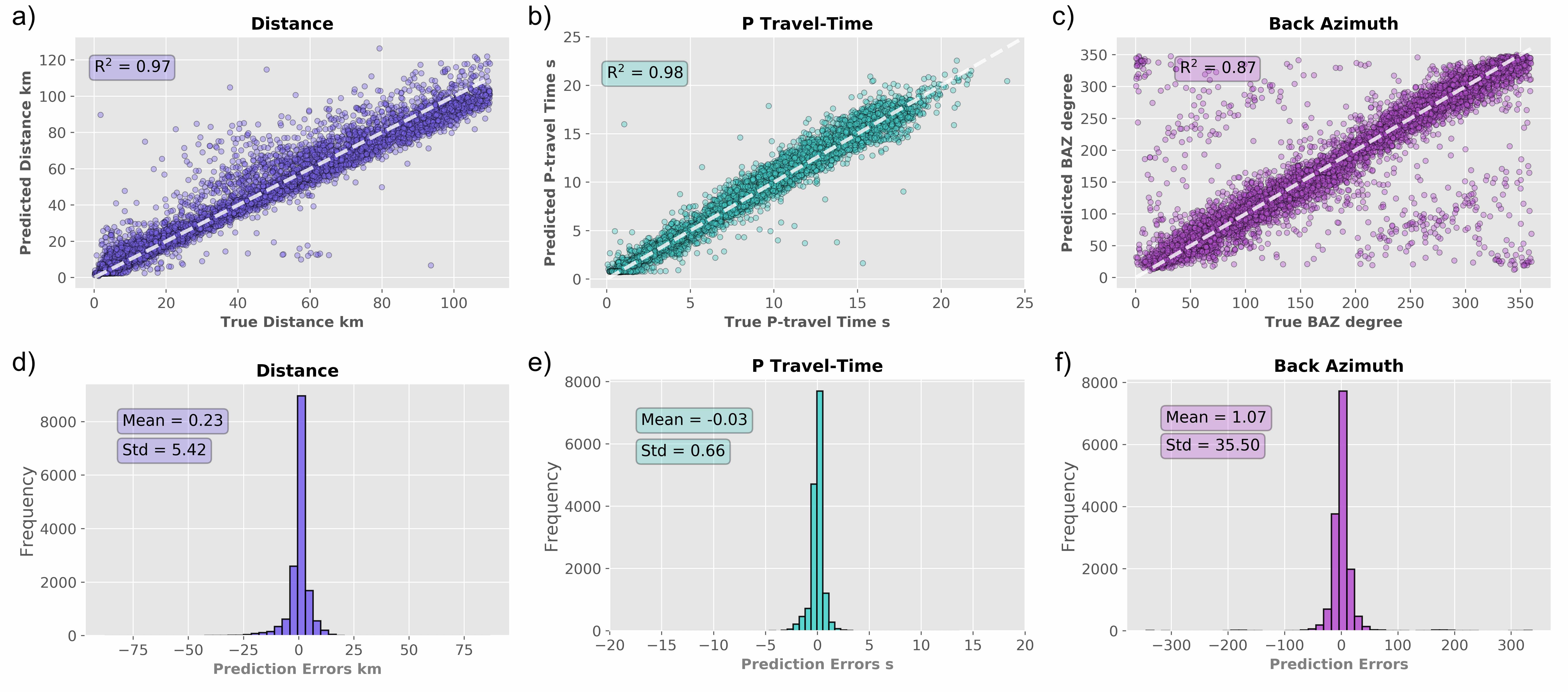}
\begin{figure*}[!htbp]
\centering \makeatletter\IfFileExists{images/78a43e04-8d95-4843-b77d-81eeffa3b4cf-ufig_5.jpg}{\includegraphics{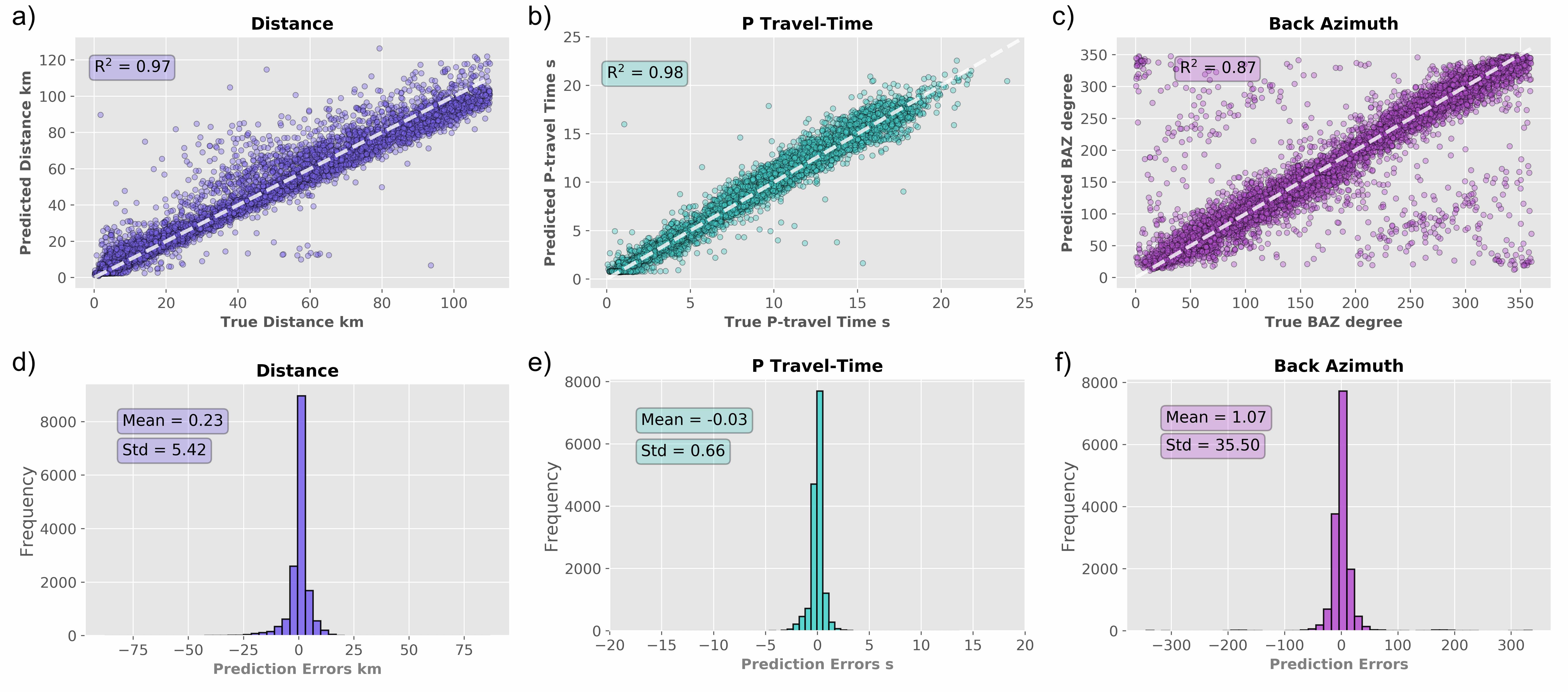}}{}
\makeatother 
\caption{{Test results for distance, P travel time, and back azimuth.}}
\label{f-5fab5bbf4c0e}
\end{figure*}
\egroup
There is some positive correlation between the estimated uncertainties and prediction errors for both distance and P travel time (Figure~\ref{f-436317dad628}), as expected.  This correlation is slightly stronger for the epicentral estimates, which suggests that estimated uncertainties might be used in practical applications to identify uncertain predictions. In both cases, the aleatory uncertainties reflect the errors better than the estimated epistemic uncertainties. This indicates a lesser role for model errors in the final output.

\bgroup
\fixFloatSize{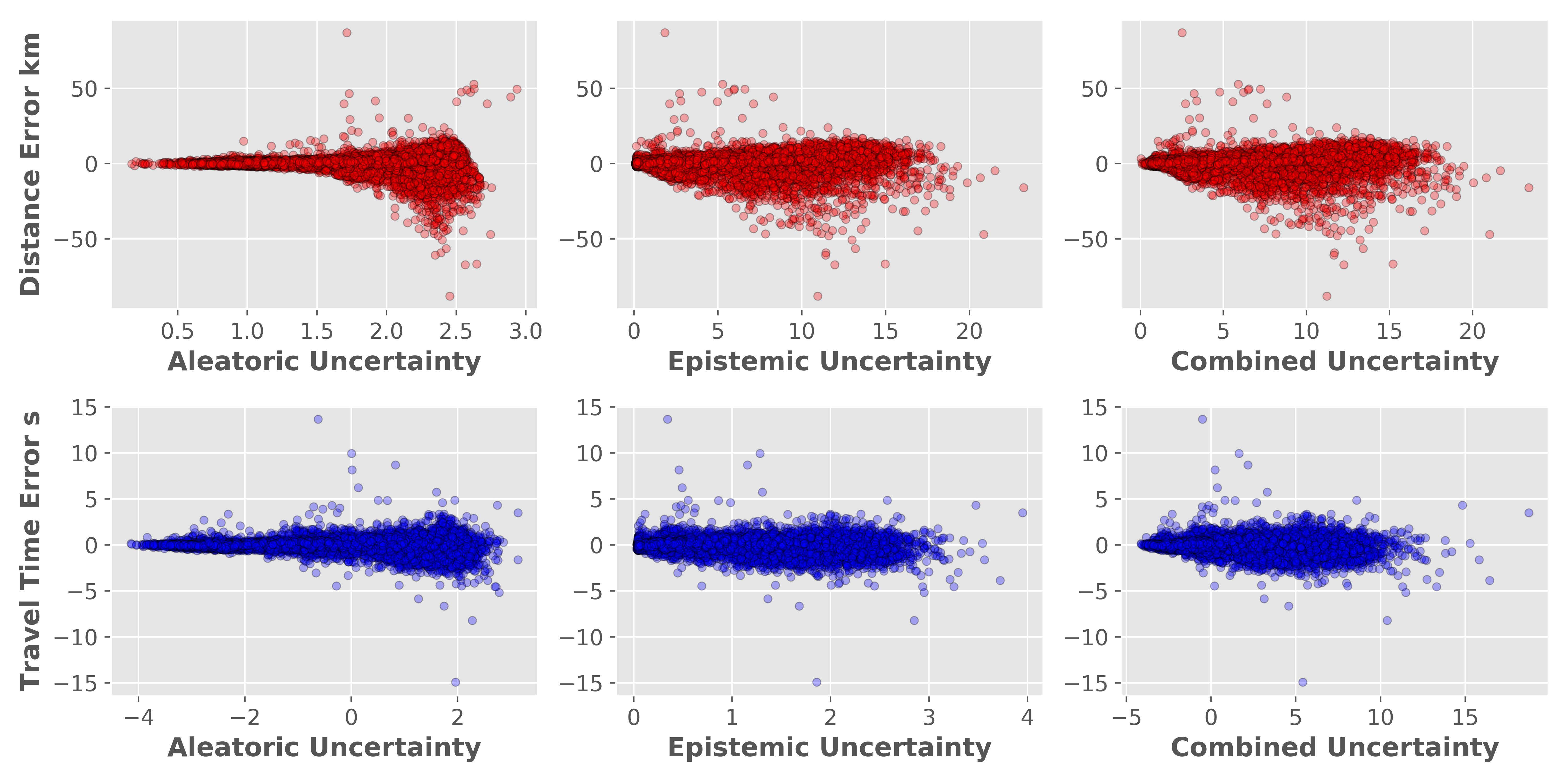}
\begin{figure*}[!htbp]
\centering \makeatletter\IfFileExists{images/6c12f4f8-8711-470e-ad40-9088a6a7b25f-ufig_6.jpg}{\includegraphics{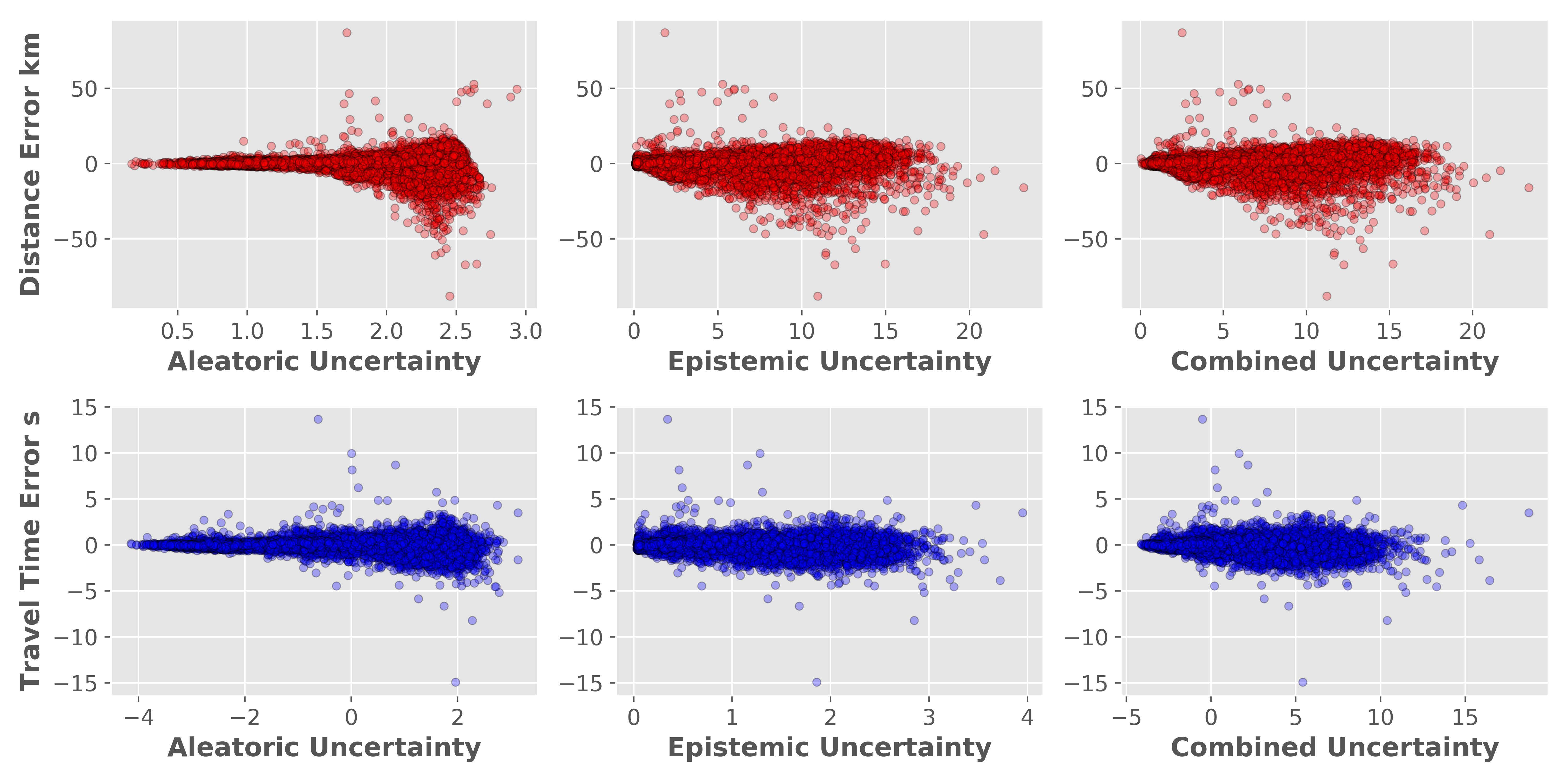}}{}
\makeatother 
\caption{{Prediction errors as a function of estimated uncertainties for epicentral distance and P travel time. }}
\label{f-436317dad628}
\end{figure*}
\egroup
We do not estimate the aleatory uncertainty for the back-azimuth estimates due to technical complications; however, we still observe a weak positive correlation between the estimated model uncertainty and the errors in the predictions (Figure~\ref{f-b77bb531c335}).

\bgroup
\fixFloatSize{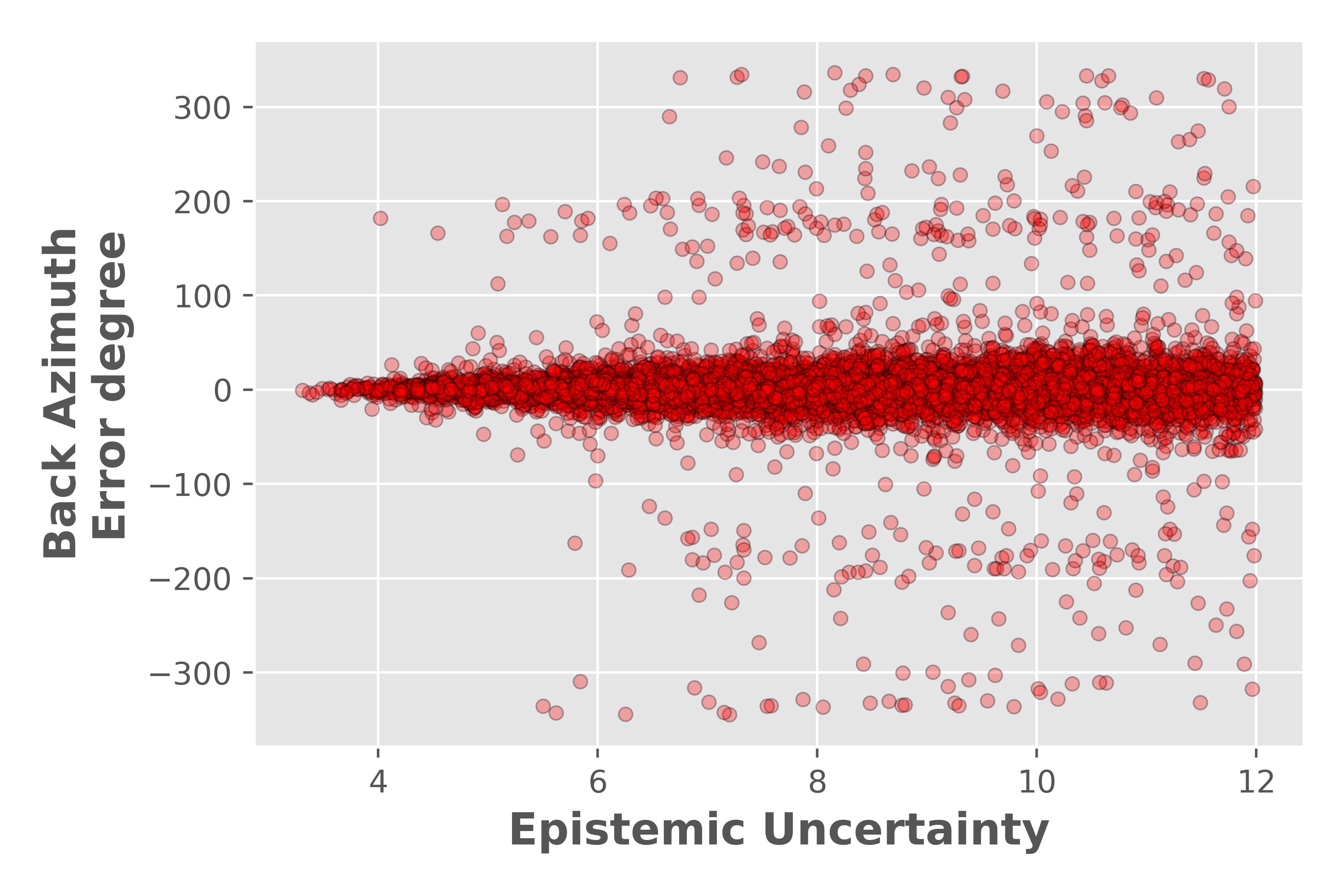}
\begin{figure}[!htbp]
\centering \makeatletter\IfFileExists{images/d13ee713-7358-4df0-856f-c18009b83e46-ufig_7.jpg}{\includegraphics{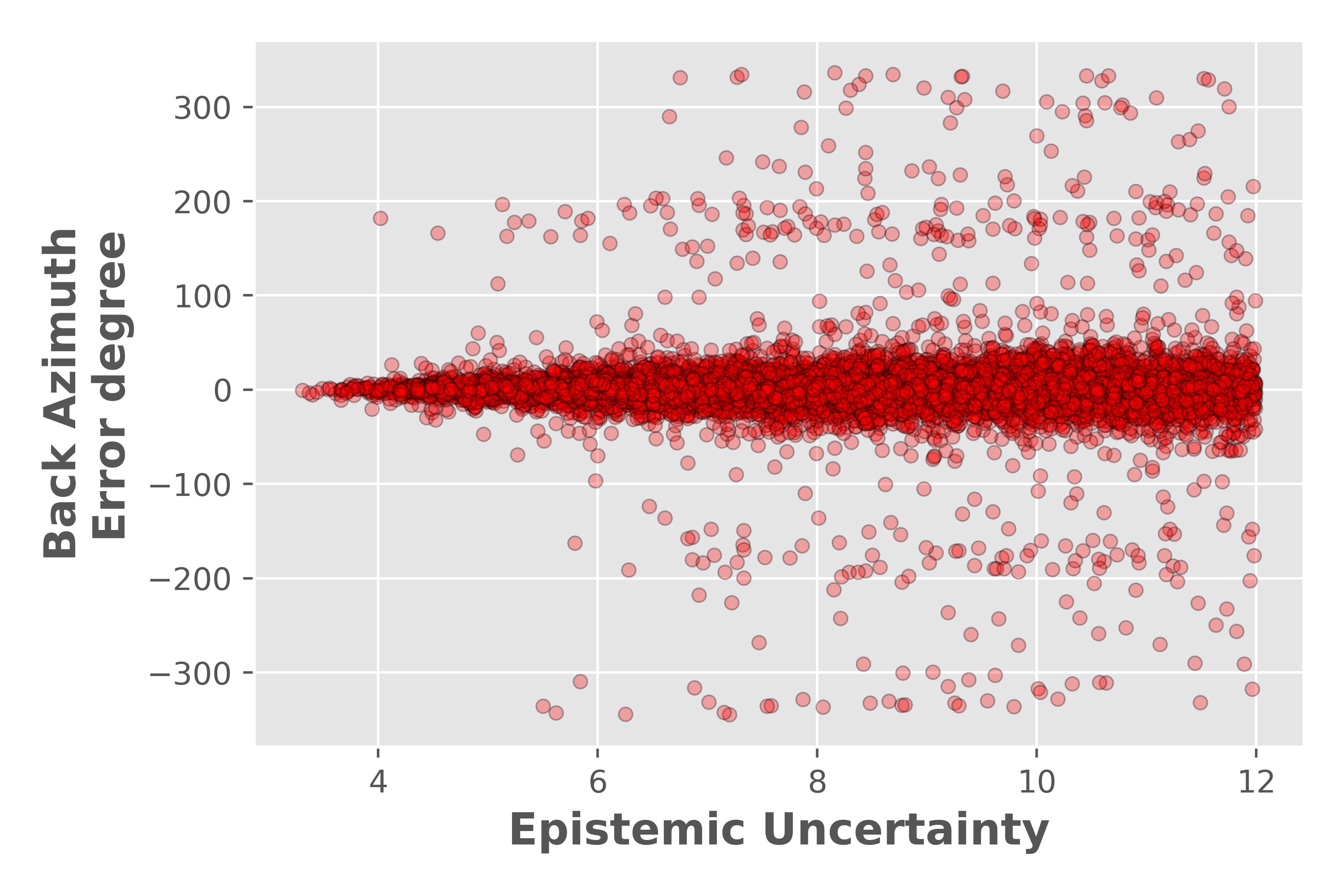}}{}
\makeatother 
\caption{{Estimated model uncertainties for test set and its relation with the prediction errors of back azimuth. }}
\label{f-b77bb531c335}
\end{figure}
\egroup
Errors in both distance and P travel time estimates seem to increase slightly with the increase in station-event distance; however, such a correlation is not apparent for the back-azimuth estimates (Figure~\ref{f-d040d4de801d}). The increase in distance has a clear effect on both estimated aleatory and epistemic uncertainties where the effect on aleatory uncertainty follows a power law, but is near-linear for the epistemic uncertainty. 

While the effects of signal-to-noise ratio of waveforms on regression errors are visible, the correlation with the estimated uncertainties is less clear. Events with magnitudes larger than 2.5 have higher uncertainties. This is likely due to the fact that we have less training data for larger magnitudes (Figure~\ref{f-d040d4de801d}).

An interesting observation is the negative bias in distance estimation for deeper events (where the predicted distances are larger than the actual distances) (Figure~\ref{f-d040d4de801d}). This explains the over estimations in the upper left side of the regression line in Figure~\ref{f-5fab5bbf4c0e}-a. Both epistemic and aleatory uncertainties are higher for the deeper events, which are fewer in number in the training dataset. We see a similar (though weaker) trend of higher epistemic uncertainties for the back-azimuth estimates, but most of the errors for the back-azimuth estimates are caused by shallow events.

\bgroup
\fixFloatSize{images/62fae2e5-0aa0-454f-8957-661d5bfbd351-ufig_8.jpg}
\begin{figure*}[!htbp]
\centering \makeatletter\IfFileExists{images/62fae2e5-0aa0-454f-8957-661d5bfbd351-ufig_8.jpg}{\includegraphics{images/62fae2e5-0aa0-454f-8957-661d5bfbd351-ufig_8.jpg}}{}
\makeatother 
\caption{{Relations between errors and uncertainties with different characteristics of events and waveform. }}
\label{f-d040d4de801d}
\end{figure*}
\egroup
\textbf{Location Results}

We used the distance and back-azimuth predictions to estimate the epicenter of the associated event for each observation in the test set. We calculate the error ellipse for each epicentral location based on estimated uncertainties for distance and back-azimuth and their projections onto the reference Earth model. We use the P travel time estimates to calculate origin times and to provide a rough estimate of earthquake depth. For the depth estimation we assume that the P waves follow a straight-line path between source and station. We assumed a velocity of 5.6 km/s for the P wave and calculated the incident angle using the estimated distance the P wave has traveled together with the estimate of epicentral distance. Estimated locations and associated error ellipses for 16 events are shown in Figures~\ref{f-6f2fa81cc7be} and~\ref{f-a51e1cdeb099}.

 We estimate locations and errors for each observation (station) and averaged for each event based on the number of available observations. To demonstrate generalization of the model, the examples are selected from different regions in Asia, Africa, Central US, Nevada, San Juan island, Southern California, and Alaska. The uncertainties for the horizontal location, depth, and origin time for the cataloged events are presented if they have been reported by the monitoring networks. We selected examples randomly to reflect an unbiased representation of the model's performance. Figure~\ref{f-6f2fa81cc7be} presents some examples with moderate errors. Errors in the origin time estimates are the lowest in general, which is consistent with the regression results. From the error ellipses, we can see a higher contribution of uncertainties in the back-azimuth estimations, which is again in agreement with the regression results. Location estimate errors are in a reasonable range considering the reported uncertainties for the catalog locations (e.g. Figure~\ref{f-6f2fa81cc7be} -f, g, and h); however, we note that our location estimates are based on only single-station observations and without the use of any velocity model. Figure~\ref{f-6f2fa81cc7be} -c, d, and e suggest that if we include the location uncertainties for individual estimates into the averaging process, the final location estimates might improve for the cases with multiple observations. 

A surprising result in these examples are the relatively good estimates of earthquake depths. Depth estimation is a persistently difficult and uncertain, yet important factor in source characterization. Our depth estimate results are within an acceptable range  considering that we do not train our network directly for end-to-end learning of depth (because of the high uncertainty and inhomogeneity in reported depths by different networks around the world that make the labeling and learning process challenging).  Instead, our estimates are derived by combining the predicted values for the P travel distance and epicentral distance.

\bgroup
\fixFloatSize{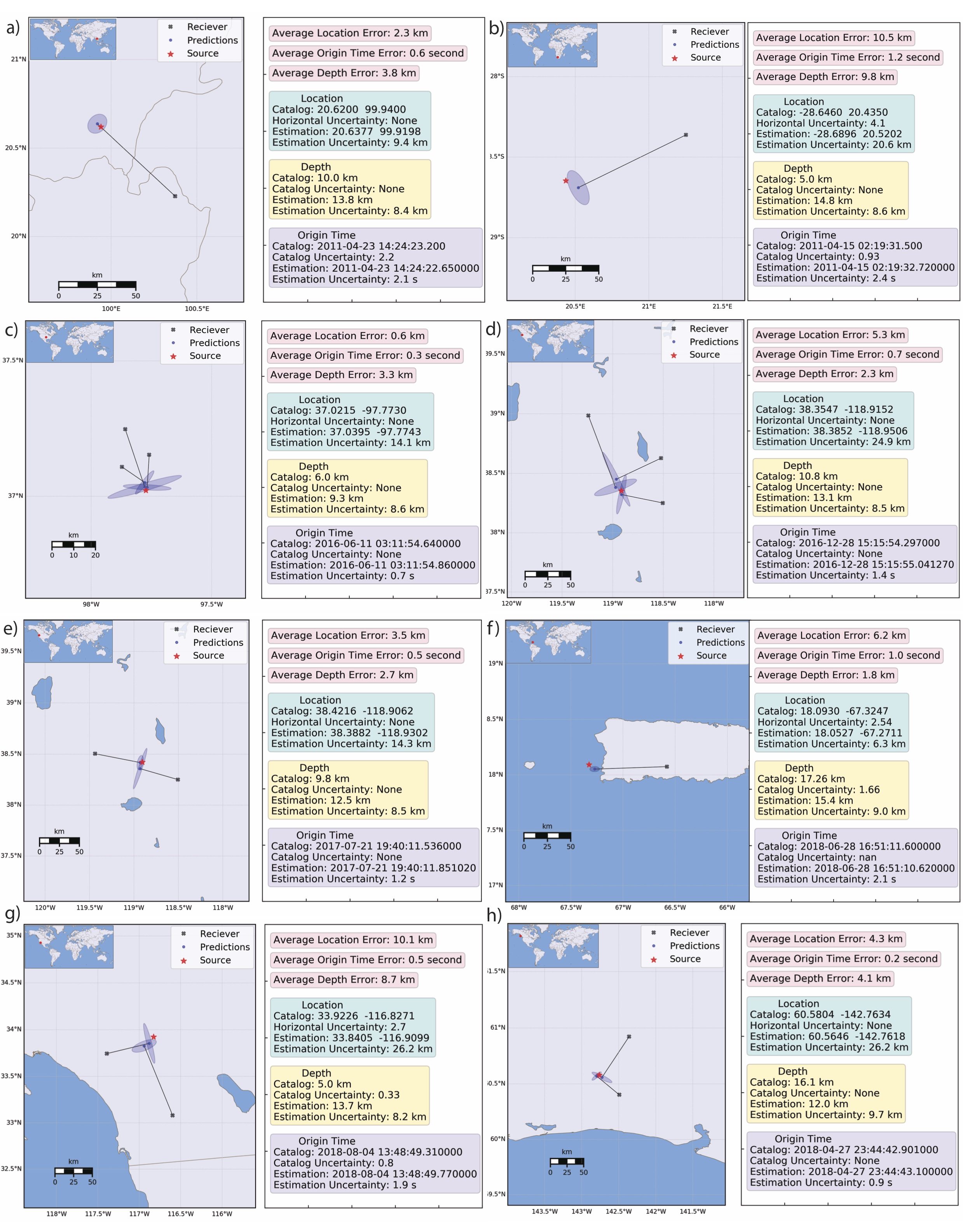}
\begin{figure*}[!htbp]
\centering \makeatletter\IfFileExists{images/19c4df76-90ab-4026-970b-e1a75cc35511-ufig_9.jpg}{\includegraphics{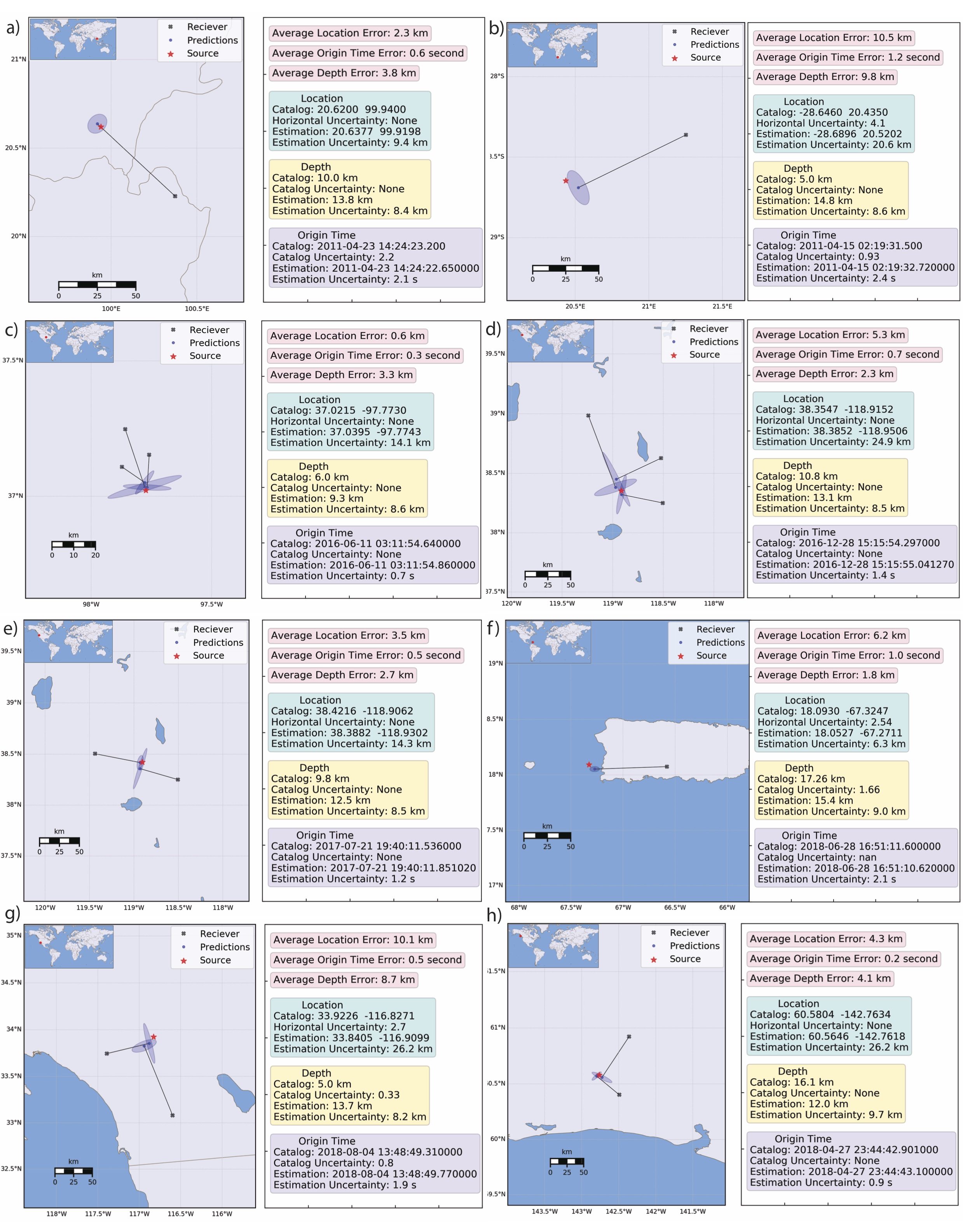}}{}
\makeatother 
\caption{{Single-station location estimates and associated error based on predicted back-azimuth and epicentral distance. For each event, we estimate errors based on averaged results (without weighting) for multiple observations. a) is an M\ensuremath{_{L}} 1.7 earthquake in Myanmar. b) is an M\ensuremath{_{L}} 1.6 in south Africa. c) an M\ensuremath{_{L}} 2.1 in Southern Kansas. d and e) are respectively an M\ensuremath{_{L}} 2.3 and M\ensuremath{_{L}} 1.3 south of Reno, Nevada. f) is an m\ensuremath{_{d}} 2.19 in San Juan island. g) is an M\ensuremath{_{L}} 3.1 northwest of Palm Springs in south California. h) is an M\ensuremath{_{L}} 1.3 in Alaska. }}
\label{f-6f2fa81cc7be}
\end{figure*}
\egroup
Figure~\ref{f-a51e1cdeb099} presents examples where our method performed poorly; however, even in the cases of unsuccessful location, single-station predictions are more-or-less pointing toward the source location. In most of these cases estimated location uncertainties are relatively large, which could be used to distinguish between these estimates and more reliable ones. An interesting observation is that even in these cases where location errors are large (mostly due to errors in the back-azimuth estimates), estimated origin time is still very close to the ground truth. This robust estimation of origin time in addition to a rough estimation of back-azimuth angle could be used for event association across a network.

\bgroup
\fixFloatSize{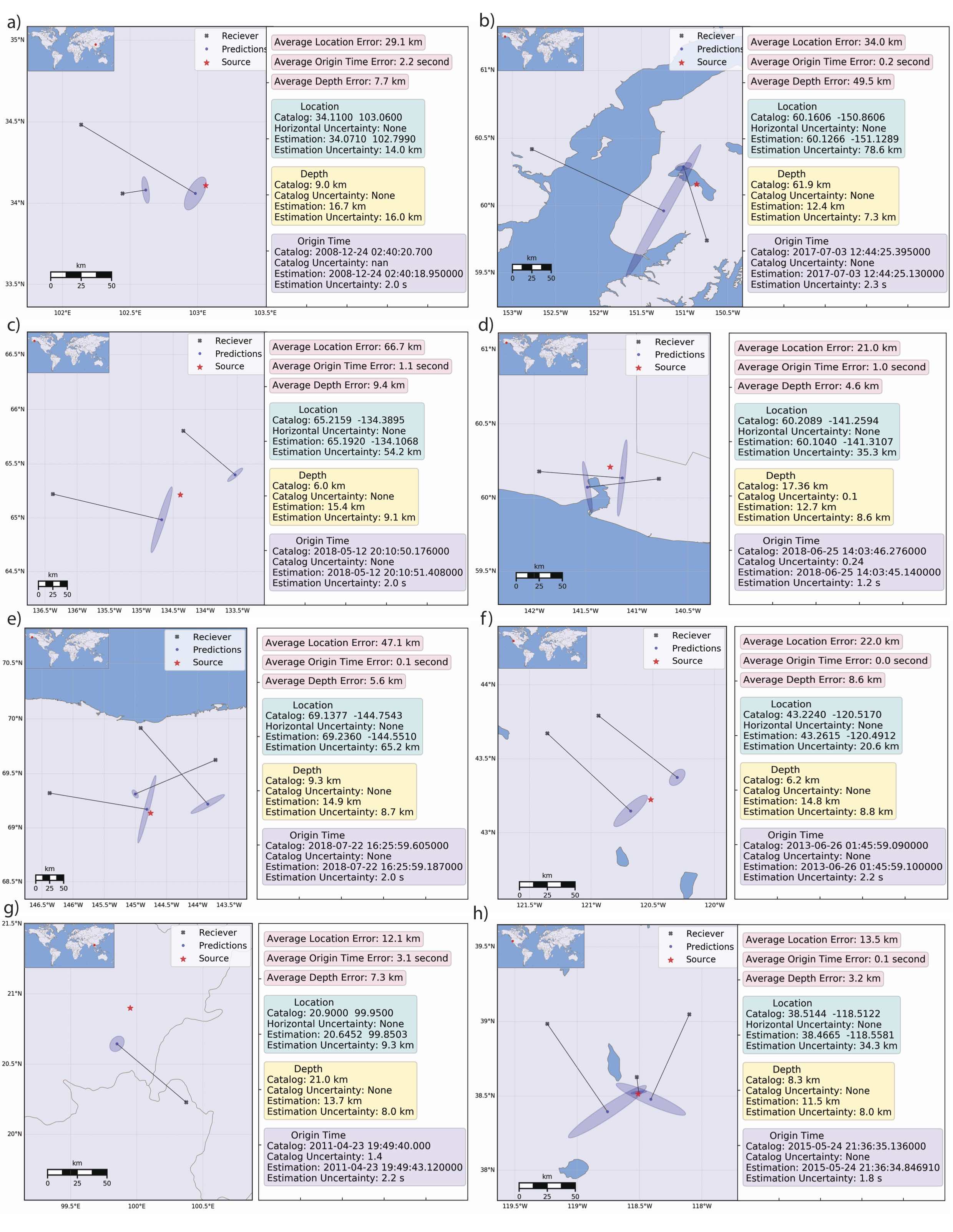}
\begin{figure*}[!htbp]
\centering \makeatletter\IfFileExists{images/d8fd99f5-5a26-40c2-85ce-e081b3c67c9d-ufig_10.jpg}{\includegraphics{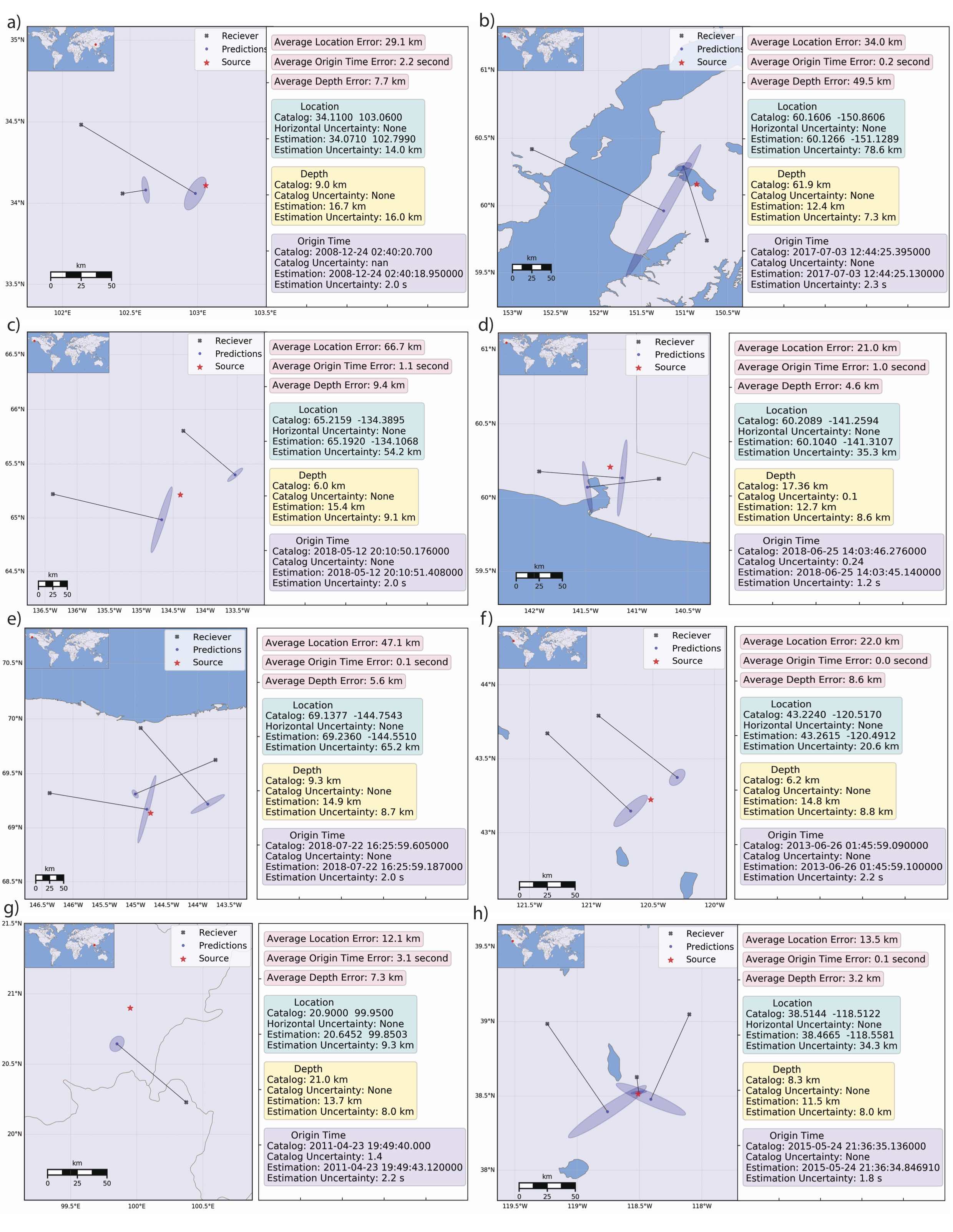}}{}
\makeatother 
\caption{{Single-station location estimates and associated error based on predicted back-azimuth and epicentral distance. For each event, errors are calculated based on averaged results (without weighting) for multiple observations. a) is an M\ensuremath{_{L}} 2.5 event in central China, b) is an M\ensuremath{_{L}} 1.6 in southern Alaska, c) is an M\ensuremath{_{L}} 2.5 in northwest Canada. d) is an M\ensuremath{_{L}} 2.1 in southeast Alaska. e) is an M\ensuremath{_{L}} 1.8 in northern Alaska. f) is an M\ensuremath{_{L}} 2.3 in Oregon, g) is an M\ensuremath{_{L}} 2.2 in Myanmar, h) is an M\ensuremath{_{L}} 1.3 in Nevada. }}
\label{f-a51e1cdeb099}
\end{figure*}
\egroup
To get a broader view of the performance of our location estimates, we plotted the predicted epicenters paired with the cataloged locations for two regions of Alaska and northern California inFigure~\ref{f-ed2735af758d}. We can see that the predicted locations reveal the overall pattern of seismicity correctly, and that the outliers are sparse. For example, the linear seismicity on the northern end of the creeping section of the San Andreas fault (between 36 and 37 degrees north) and the event cluster in Geysers (38.8 degrees north) is clearly recovered in (Figure~\ref{f-ed2735af758d}-b). These are averaged results for the event-based estimates without removing any high-uncertainty estimates or any weighting of the results during the averaging.

\bgroup
\fixFloatSize{images/0b78e05d-7456-411e-b77a-311826420a20-ufig_11.jpg}
\begin{figure*}[!htbp]
\centering \makeatletter\IfFileExists{images/0b78e05d-7456-411e-b77a-311826420a20-ufig_11.jpg}{\includegraphics{images/0b78e05d-7456-411e-b77a-311826420a20-ufig_11.jpg}}{}
\makeatother 
\caption{{Single-station location predictions connected to their associated ground truth (locations in the catalogs) for Alaska (a) and Northern California (b). }}
\label{f-ed2735af758d}
\end{figure*}
\egroup
The overall performance of our locations for the entire test set can be seen from the error distributions presented in Figure~\ref{f-328a31aa8989}. Our method predicts epicenter, origin time, and depth with a mean error of 7.3 km, 0.4 seconds, and 6.7 km respectively. These are in agreement with our previous observations and with regression results.

\bgroup
\fixFloatSize{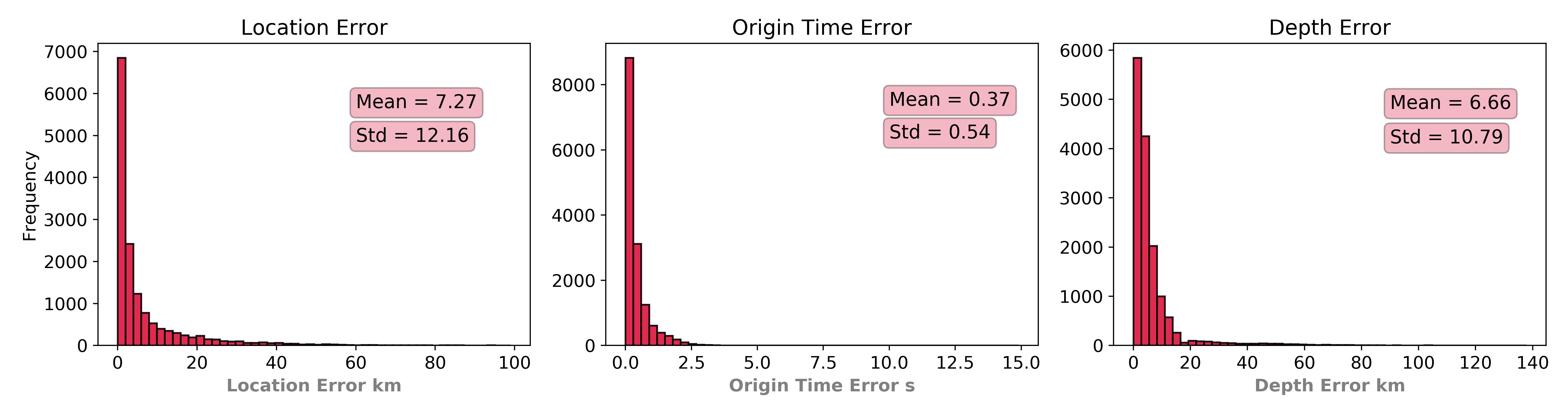}
\begin{figure*}[!htbp]
\centering \makeatletter\IfFileExists{images/0943ee8c-d896-471d-93bb-8c652a7cea0e-ufig_12.jpg}{\includegraphics{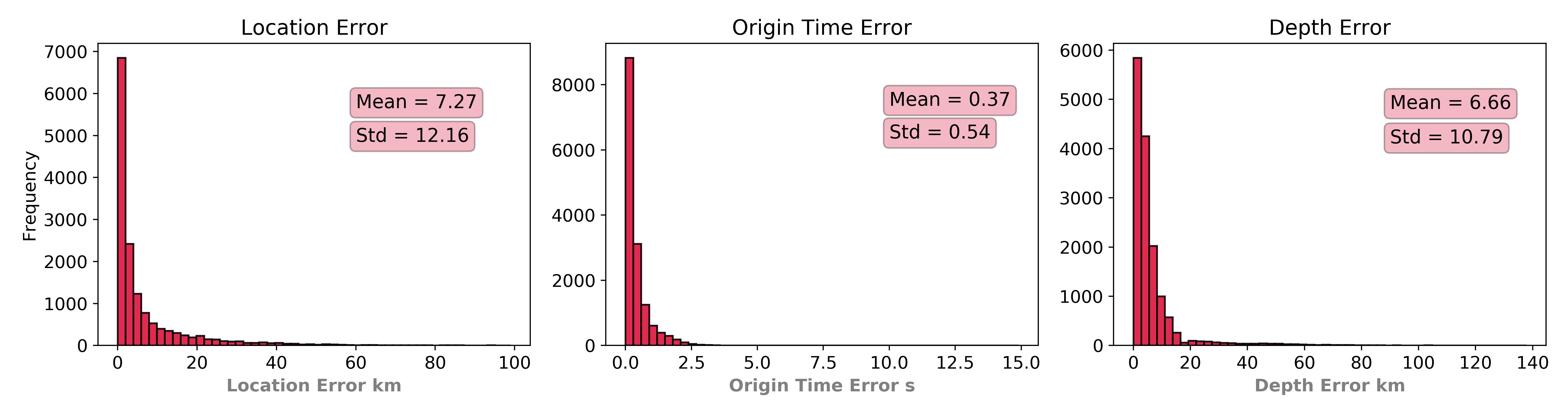}}{}
\makeatother 
\caption{{Statistics of location, origin time, and depth estimations for all the data (globally distributed) in the test set. Results are based on single-station estimates averaged for each event. }}
\label{f-328a31aa8989}
\end{figure*}
\egroup
To understand these errors and their potential sources further, we show each of them plotted as a function of event magnitude, depth, predicted uncertainties, and reported uncertainty in the catalog in Figure~\ref{f-2dcf73a550a4}. Larger events tend to have larger prediction errors, which may be attributable to the more sparse training data for larger events in the dataset. The strong performance for small events indicates the sensitivity of the method. The ability to locate smaller events is important because small events are exactly those that are most likely to be detected on fewer stations, and a single-station method might be the only way to locate them. There is a strong correlation between errors in the depth estimation and the event depth. Very shallow events and events deeper than 20 km have larger errors. Predicted uncertainties seem to correlate with the errors in the predictions. The fifth row in Figure~\ref{f-2dcf73a550a4} suggests some part of the mismatch between the predictions and ground truth might be due to uncertainties in the reported location, origin time, or depth. The relations between estimated uncertainties (by our model) and reported uncertainties in the earthquake catalogs are presented in the last column.

\bgroup
\fixFloatSize{images/99abd9c1-2deb-439d-9a9d-f9830ab82b92-ufig_13.jpeg}
\begin{figure*}[!htbp]
\centering \makeatletter\IfFileExists{images/99abd9c1-2deb-439d-9a9d-f9830ab82b92-ufig_13.jpeg}{\includegraphics{images/99abd9c1-2deb-439d-9a9d-f9830ab82b92-ufig_13.jpeg}}{}
\makeatother 
\caption{{The relationships between prediction errors and event magnitude, depth, predicted uncertainties, and reported uncertainty in the catalog. }}
\label{f-2dcf73a550a4}
\end{figure*}
\egroup

\section{Discussion and Conclusions}
We present a successful application of deep learning for earthquake location based on single-station observations. The model is trained and tested using a global data set. Our test results indicate that our neural network can directly learn a general mapping function between the raw 3-component seismograms (and known P and S arrivals) and epicentral distance, P travel time, and back azimuth without the need for a local velocity model. Distance and back-azimuth predictions can be directly expanded to regional and teleseismic distances by training with waveforms from these events; however, the current approximate depth estimation procedure will not be applicable due to a combination of more complex wave propagation and the sphericity of the Earth. A distinct advantage of our approach lies in its Bayesian framework, which provides an estimate of uncertainties in data and model and allows us to estimate confidence intervals in the final estimated location. 

Our strategy in using a multi-task network for closely related tasks (i.e., distance and travel time) and separate networks for the other task (i.e., back-azimuth) improve the overall performance. The designed networks are light with a relatively low number of trainable parameters, which requires fewer data in the training set. Availability of larger training sets with high-quality labels will allow us to design more flexible/powerful networks and potentially to improve the performance by reducing epistemic uncertainties. Deeper networks with more convolutional layers have the potential, for example, to reduce the sensitivity of the model to noise. 

Building large training sets for these tasks is challenging because, in addition to the quality of general labeling such as existence of earthquake signal in the window, accuracy of picks, accuracy of metadata used for estimate of back-azimuth, P travel time, and distance etc, station orientations are also important. Our current method does not consider station orientations that differ from the geographical orientations, and is not applicable to the borehole stations. This prevents us from using the high signal-to-noise ratio data provided by borehole instruments. 

The largest uncertainty in location results is caused by uncertainties in the back-azimuth. Errors and high uncertainties in back-azimuth estimates may be due to seismic station installation and orientation error \unskip~\cite{542319:14177663,542319:14177664}; however, learning the orientation angle using neural networks is technically challenging as well.  Utilizing more advanced methods for continuous orientation estimation (e.g. \unskip~\cite{542319:14404347}) might improve the results. Moreover, previous studies \unskip~\cite{542319:14404468} showed the length of the window used for polarization estimation play an important role in precision of back-azimuth estimation. Optimizing this window length based on dominate signal frequency might be a potential solution for this\unskip~\cite{542319:14404511}.  We note that our method might provide a novel approach to detecting mis-oriented seismometers.

 Our results indicate superior performance of the neural-network based approach compared with traditional singel-station location methods (e.g. \unskip~\cite{542319:14404592,542319:14404634,542319:14404635}). The proposed method can provide a rapid estimate of earthquake location directly from single instruments. This may be useful for rapid public reporting or earthquake early warning systems\unskip~\cite{542319:12368995}. Lockman and Allen\unskip~\cite{542319:14404645} investigated the accuracy of event parameter determination using single station for the purpose of early warning using P-wave arrival only. They concluded that estimated hypocentral distance and backazimuth with accuracy of$\pm15 $ km and $\pm20^\circ $ respectively that was obtained by the high-quality stations in southern California are sufficient to provide useful early warning. We showed that our approach can result in a much higher accuracy for smaller events with more complicated high-frequency waveforms. Our approach has the potential to function even using a portion of waveform (e.g. P-wave only) through augmentation. On the other hand, the provided information about P and S arrival times (as an extra vector in addition to the three-component waveforms) are mainly used to speed up the training process by directing the attention of the network to the informative part of the input data and our tests indicate the network can perform well even when picks are not available.

\section{Methods}
\textbf{Bayesian deep learning}

Bayesian deep learning lies at the intersection of Bayesian statistical theory and deep learning, and makes it possible to express different aspects of uncertainty in deep learning models probabilistically. In Bayesian-deep learning, all quantitities are represented as probability distributions rather than a point estimates. Prior probability distributions are applied over model parameters and are used to represent how they relate to the data. Uncertainties in the observed data are then inferred using probability theory. Learning from a training dataset is done by transforming the prior probability distributions (defined before training on data), into posterior distributions (determined after observing data) that captures a set of plausible model parameters given the data\unskip~\cite{542319:12461547}.

\textbf{Types of Uncertainties}

Three types of uncertainty contribute to the model predictions; aleatory uncertainty, epistemic uncertainty, and ontological uncertainty. 

The aleatory (also known as irreducible, inherent, stochastic, or type-A) uncertainty captures uncertainty with respect to information that our data cannot explain. Aleatory uncertainty can be divided into two sub-categories of Heteroscedastic (data-dependent) and Homoscedastic (task-dependent) uncertainties. 

\textbf{Uncertainty Estimation}

Let $\widehat y $ be the output of a neural network model with a loss function $\smallmathcal l(.,.) $ (i.e. the Euclidean for regression). We denote by $\boldsymbol W $ the network's weight (model parameters) and by  $\boldsymbol X=\{x_1,\;...,\;x_N\} $ and $\boldsymbol Y=\{y_1,\;...,\;y_N\} $ a set of $N $inputs and outputs respectively. The objective is to learn unknown parameters, $\boldsymbol W $, by minimizing the loss between predictions and actual values ($\smallmathcal l(y_i,{\widehat y}_i) $). Assuming a Gaussian distribution for $\boldsymbol Y $the data likelihood is defined as$p(\boldsymbol Y\vert\boldsymbol X,\boldsymbol W) $. Given $\boldsymbol X $and $\boldsymbol Y $, the posterior probability distribution of weight, $p(\boldsymbol W\vert\boldsymbol X,\boldsymbol Y)\;=\;p(\boldsymbol Y\vert\boldsymbol X,\boldsymbol W)p(\boldsymbol W)/p(\boldsymbol Y\vert\boldsymbol X) $, can be inferred via Bayesian theory that provides a set of model parameters. The prior probability distribution over weight, $p(\boldsymbol W) $, then can be used to estimate the weight uncertainty\unskip~\cite{542319:12461322}. 

Although conceptually simple, this is difficult to perform in practice because the marginal probability, $p(\boldsymbol Y\vert\boldsymbol X) $ , can not be evaluated analytically and requires some approximations\unskip~\cite{542319:12463129,542319:12463130,542319:12463131}. In these approximation techniques, a simple distribution, $q(\boldsymbol W) $, over the network's weights is learned by minimizing the Kullback-Leibler (KL) divergence between the approximating distribution and the full posterior, $KL(q(\boldsymbol W)\vert\vert p(\boldsymbol W\vert\boldsymbol X,\boldsymbol Y)) $.

Here, we use Monte Carlo dropout sampling\unskip~\cite{542319:12459951,542319:12463131} as a variational Bayesian approximation for this purpose. Dropout\unskip~\cite{542319:12370981}, is a simple regularization technique commonly used to prevent over-fitting in deep learning. Dropout randomly removes network units during the training and by doing this samples from a number of trained networks with reduced width. For test data, dropout approximates the effect of averaging the predictions of thinned networks using the weights of the un-thinned network\unskip~\cite{542319:12464545}. Gal and Ghahramani\unskip~\cite{542319:12459951} showed that a neural network with with dropout applied during both training and test times is mathematically equivalent to an approximation to the probabilistic deep Gaussian process\unskip~\cite{542319:12480069}.  This is equivalent to Monte Carlo sampling from an approximate posterior distribution over models that find an approximating distribution ($q(\boldsymbol W) $) with minimum KL divergence to the posterior probability distribution.  This technique is computationally efficient, and unlike other approximations, can be easily applied to large and complex networks. 

Here, we use dropout to approximate the posterior distribution for a model, $f $, that maps an input directly to the predictive mean and variance:
\let\saveeqnno\theequation
\let\savefrac\frac
\def\dispfrac{\displaystyle\savefrac}
\begin{eqnarray}
\let\frac\dispfrac
\gdef\theequation{1}
\let\theHequation\theequation
\label{dfg-4990e262fcae}
\begin{array}{@{}l}\left[\widehat y,\widehat\sigma ^{2}\right]=f(x)\end{array}
\end{eqnarray}
\global\let\theequation\saveeqnno
\addtocounter{equation}{-1}\ignorespaces 
where $f $ is a Bayesian convolutional neural network that transform the input seismogram, $x $, directly to an output value, $\widehat y\;\in\mathbb{R} $, and a measure of aleatory uncertainty given by variance, $\sigma ^{2} $. This variance (noise in the data) can be directly estimated from the data using a customized loss function that includes both the regression and an uncertainty regularization term\unskip~\cite{542319:12459953}:
\let\saveeqnno\theequation
\let\savefrac\frac
\def\dispfrac{\displaystyle\savefrac}
\begin{eqnarray}
\let\frac\dispfrac
\gdef\theequation{2}
\let\theHequation\theequation
\label{dfg-426f090e9702}
\begin{array}{@{}l}{\smallmathcal l}_{BNN}\;=\;\frac1{2N}\sum_{i=1}^{N}exp(-s_i)\;\left|\left|y_i-{\widehat y}_i\right|\right|^{2}+\frac12s_i\end{array}
\end{eqnarray}
\global\let\theequation\saveeqnno
\addtocounter{equation}{-1}\ignorespaces 
where $s_i=\log\;{\widehat\sigma }_i^{2} $. The variance is data-dependent and is learned implicitly from the loss function during the training without the need for uncertainty labels. It represents the hetroscedastic aleatoric uncertainties and is useful in cases where input data point have different noise levels. The predictive uncertainty (a combination of epistemic and aleatory uncertainty) then can be estimated as:
\let\saveeqnno\theequation
\let\savefrac\frac
\def\dispfrac{\displaystyle\savefrac}
\begin{eqnarray}
\let\frac\dispfrac
\gdef\theequation{3}
\let\theHequation\theequation
\label{dfg-6cc13a440796}
\begin{array}{@{}l}Var(y)\approx\frac1T\sum_{t=1}^{T}\widehat y_t^{2}-\left(\frac1T\sum_{t=1}^{T}{\widehat y}_t\right)^{2}+\frac1T\sum_{t=1}^{T}\widehat\sigma _t^{2}\end{array}
\end{eqnarray}
\global\let\theequation\saveeqnno
\addtocounter{equation}{-1}\ignorespaces 
where $T $ is the number of Monte Carlo dropout samples. 

Simultaneous prediction of output and uncertainty by network makes the model robust to noisy data. The loss function acts as an intelligent regression function that allows the network to learn to attenuate the effect of erroneous labels by adapting the residual's weighting\unskip~\cite{542319:12459953}. 

\textbf{Temporal Convolution Networks}

Earthquake signals contain sequential information.  High-frequency compressional P-waves arrive before transverse S-waves, which in turn arrive before dispersive surface waves. These temporal dependencies among different components of an earthquake signal arise from the physics of elastic wave propagation and motivate the use of neural networks that are capable of sequence modeling.  For this reason, recurrent neural nets (e.g. LSTM\unskip~\cite{542319:14121919} or GRU\unskip~\cite{542319:14121961}) particularly suitable for modeling earthquake signals. 

It has been shown that a combination of LSTMs and CNNs can achieve good performance in learning both the local structures in a seismic signal and the temporal dependencies among these structures while reducing the computational time of sequential learning\unskip~\cite{542319:14121518}; however, recent studies have also demonstrated that certain convolutional architectures can outperform recurrent architectures across a diverse range of sequential-modeling tasks and data sets, while demonstrating the same capacity with longer effective memory\unskip~\cite{542319:12369449}. These convolutional architectures have the advantages of architectural simplicity, parallelism, flexible receptive field size, stable gradients, low memory requirements for training, and variable length inputs. Such temporal convolution networks, are a family of autoregressive feed-forward models with causal dilated convolutions and residual connections.  Temporal convolution networks are the latest development in sequential modeling in speech recognition, time series analysis, natural language processing, and signal processing. 

Our network for distance/P-travel time estimation consists of 1D convolutional layers where convolutions are causal and dilated. Causal convolution means that the information flow is only from past to future, such that to make a prediction at time point \textit{t}, only the information up to time \textit{t}, is used. 

Dilated convolutions (also called trous, or convolution with holes) are used to learn longer history by increasing the receptive field exponentially \unskip~\cite{542319:12370832,542319:12370874}. They allow for having very large receptive fields and they can learn multi-scale structures without greatly increasing the number of parameters. A large receptive field with small kernel size makes it possible to combine local and global informationFigure~\ref{f-13007899a493}. 

Residual structure\unskip~\cite{542319:12370984}, allows a deeper network without degradation with a larger receptive field. Each residual block consists of two layers of dilated causal convolution and rectified linear units\unskip~\cite{542319:12370939}. Here we use 1D-spatial droupouts\unskip~\cite{542319:12370981} for regularization and also uncertainty estimation. Since in our architecture the input and output have different widths, a  $1\times1 $ convolution is used instead of identity mapping to match the size for addition operation.

\bgroup
\fixFloatSize{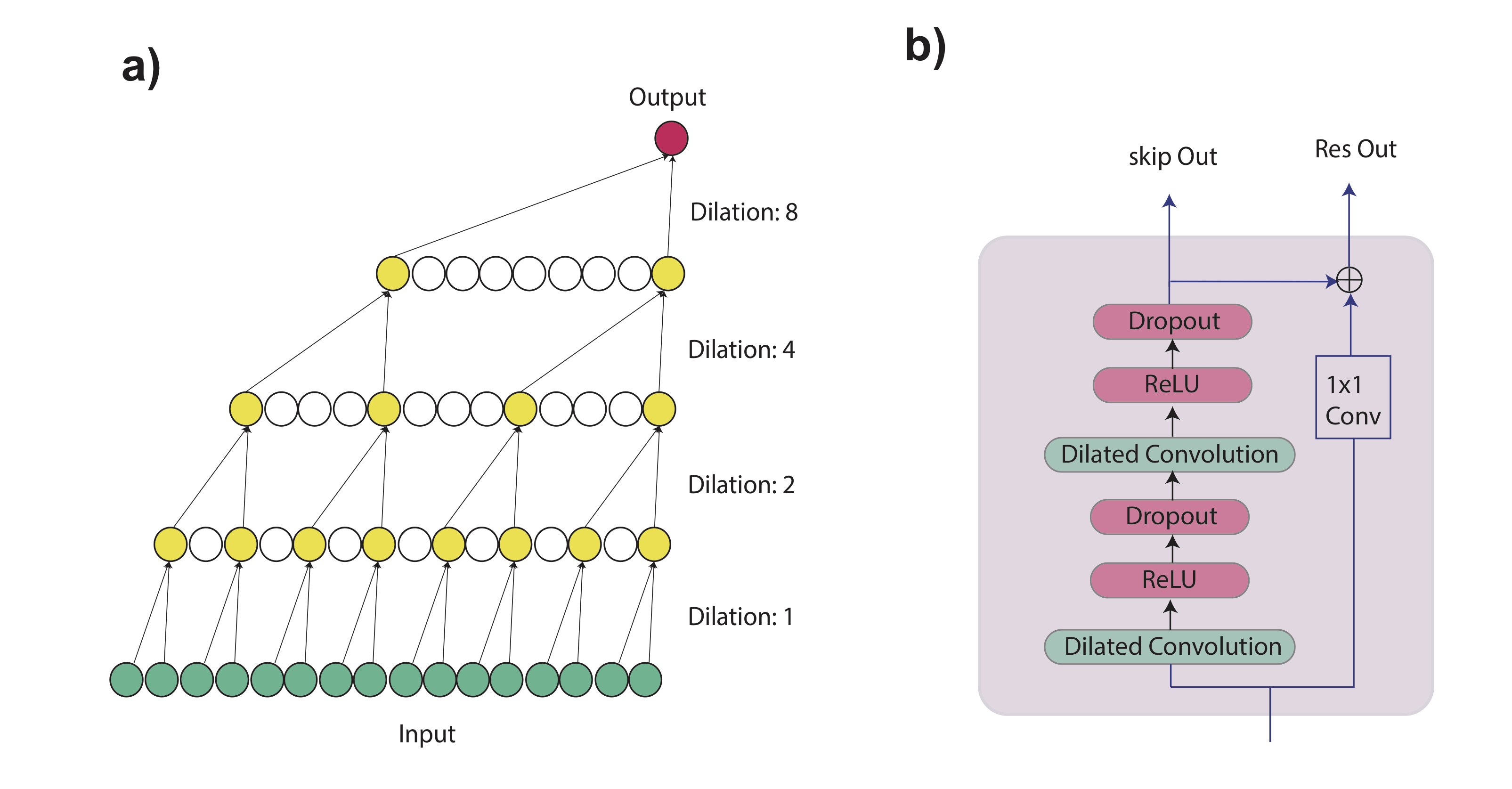}
\begin{figure*}[!htbp]
\centering \makeatletter\IfFileExists{images/223bfa7e-f68c-4f75-8e0f-a8776b967d84-ufig_14.jpg}{\includegraphics{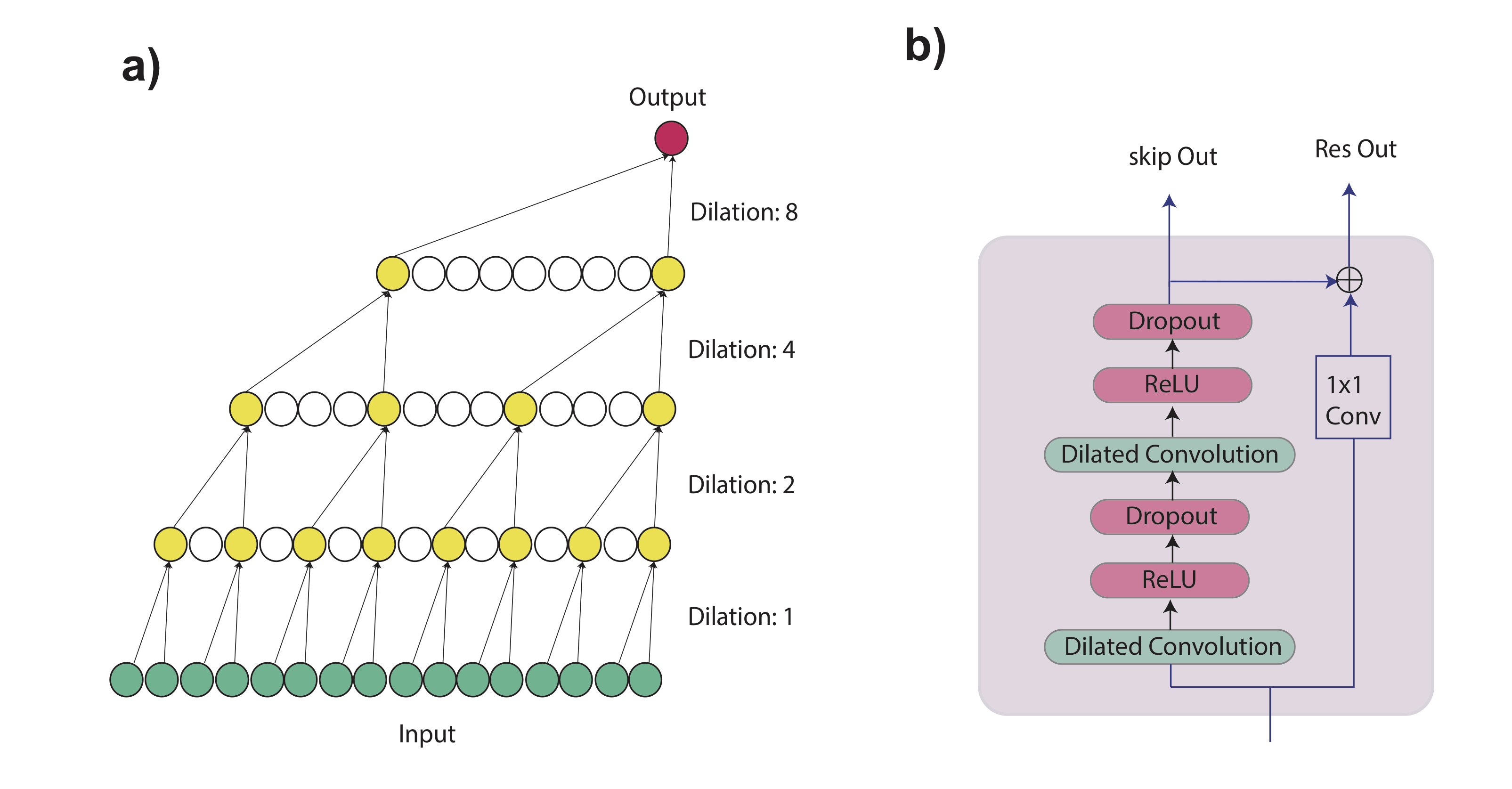}}{}
\makeatother 
\caption{{a) demonstration of dilational convolution operation for dilations 1, 2, 4, and 8. b) residual structure used over the dilational network.}}
\label{f-13007899a493}
\end{figure*}
\egroup
The dilation factor, \textit{d} is increased exponentially with the depth of the network to ensure that all of the inputs within the receptive field are convolved with some filters while allowing for a very large effective history.

\bibliographystyle{naturemag}

\bibliography{\jobname}\section*{Acknowledgments}S.M.M. was partially supported by Stanford Center for Induced and Triggered Seismicity during this project. G.C.B. was supported by AFRL under contract number FA9453-19-C-0073. 

\begin{thebibliography}{10}
\expandafter\ifx\csname url\endcsname\relax
  \def\url#1{\texttt{#1}}\fi
\expandafter\ifx\csname urlprefix\endcsname\relax\def\urlprefix{URL }\fi
\providecommand{\bibinfo}[2]{#2}
\providecommand{\eprint}[2][]{\url{#2}}

\bibitem{542319:14121517}
\bibinfo{author}{Perol, T.}, \bibinfo{author}{Gharbi, M.} \&
  \bibinfo{author}{Denolle, M.}
\newblock \bibinfo{title}{{Convolutional neural network for earthquake
  detection and location}}.
\newblock \emph{\bibinfo{journal}{{Science Advances}}}
  \textbf{\bibinfo{volume}{4}}, \bibinfo{pages}{e1700578}
  (\bibinfo{year}{2018}).

\bibitem{542319:14121518}
\bibinfo{author}{Mousavi, S.~M.}, \bibinfo{author}{Zhu, W.},
  \bibinfo{author}{Sheng, Y.} \& \bibinfo{author}{Beroza, G.~C.}
\newblock \bibinfo{title}{{CRED: A deep residual network of convolutional and
  recurrent units for earthquake signal detection}}.
\newblock \emph{\bibinfo{journal}{{Scientific reports}}}
  \textbf{\bibinfo{volume}{9}}, \bibinfo{pages}{10267} (\bibinfo{year}{2019}).

\bibitem{542319:14121519}
\bibinfo{author}{Zhu, W.} \& \bibinfo{author}{Beroza, G.~C.}
\newblock \bibinfo{title}{{PhaseNet: a deep-neural-network-based seismic
  arrival-time picking method}}.
\newblock \emph{\bibinfo{journal}{{Geophysical Journal International}}}
  \textbf{\bibinfo{volume}{216}}, \bibinfo{pages}{261--273}
  (\bibinfo{year}{2018}).

\bibitem{542319:14121520}
\bibinfo{author}{Pardo, E.}, \bibinfo{author}{Garfias, C.} \&
  \bibinfo{author}{Malpica, N.}
\newblock \bibinfo{title}{{Seismic Phase Picking Using Convolutional
  Networks}}.
\newblock \emph{\bibinfo{journal}{{IEEE Transactions on Geoscience and Remote
  Sensing}}}  (\bibinfo{year}{2019}).

\bibitem{542319:14121521}
\bibinfo{author}{Ross, Z.~E.}, \bibinfo{author}{Meier, M.-A.},
  \bibinfo{author}{Hauksson, E.} \& \bibinfo{author}{Heaton, T.~H.}
\newblock \bibinfo{title}{{Generalized seismic phase detection with deep
  learning}}.
\newblock \emph{\bibinfo{journal}{{Bulletin of the Seismological Society of
  America}}} \textbf{\bibinfo{volume}{108}}, \bibinfo{pages}{2894--2901}
  (\bibinfo{year}{2018}).

\bibitem{542319:14121522}
\bibinfo{author}{Chen, Y.}
\newblock \bibinfo{title}{{Automatic microseismic event picking via
  unsupervised machine learning}}.
\newblock \emph{\bibinfo{journal}{{Geophysical Journal International}}}
  \textbf{\bibinfo{volume}{212}}, \bibinfo{pages}{88--102}
  (\bibinfo{year}{2017}).

\bibitem{542319:14121524}
\bibinfo{author}{Ross, Z.~E.}, \bibinfo{author}{Meier, M.-A.} \&
  \bibinfo{author}{Hauksson, E.}
\newblock \bibinfo{title}{{P wave arrival picking and first-motion polarity
  determination with deep learning}}.
\newblock \emph{\bibinfo{journal}{{Journal of Geophysical Research: Solid
  Earth}}} \textbf{\bibinfo{volume}{123}}, \bibinfo{pages}{5120--5129}
  (\bibinfo{year}{2018}).

\bibitem{542319:14121525}
\bibinfo{author}{Mousavi, S.~M.}, \bibinfo{author}{Zhu, W.},
  \bibinfo{author}{Ellsworth, W.} \& \bibinfo{author}{Beroza, G.}
\newblock \bibinfo{title}{{Unsupervised Clustering of Seismic Signals Using
  Deep Convolutional Autoencoders}}.
\newblock \emph{\bibinfo{journal}{{IEEE Geoscience and Remote Sensing
  Letters}}}  (\bibinfo{year}{2019}).

\bibitem{542319:14121526}
\bibinfo{author}{Zhu, W.}, \bibinfo{author}{Mousavi, S.~M.} \&
  \bibinfo{author}{Beroza, G.~C.}
\newblock \bibinfo{title}{{Seismic signal denoising and decomposition using
  deep neural networks}}.
\newblock \emph{\bibinfo{journal}{{arXiv preprint arXiv:1811.02695}}}
  (\bibinfo{year}{2018}).

\bibitem{542319:14121527}
\bibinfo{author}{Zhang, C.}, \bibinfo{author}{van~der Baan, M.} \&
  \bibinfo{author}{Chen, T.}
\newblock \bibinfo{title}{{Unsupervised dictionary learning for signal-to-noise
  ratio enhancement of array data}}.
\newblock \emph{\bibinfo{journal}{{Seismological Research Letters}}}
  \textbf{\bibinfo{volume}{90}}, \bibinfo{pages}{573--580}
  (\bibinfo{year}{2018}).

\bibitem{542319:14121528}
\bibinfo{author}{Mousavi, S.~M.}, \bibinfo{author}{Horton, S.~P.},
  \bibinfo{author}{Langston, C.~A.} \& \bibinfo{author}{Samei, B.}
\newblock \bibinfo{title}{{Seismic features and automatic discrimination of
  deep and shallow induced-microearthquakes using neural network and logistic
  regression}}.
\newblock \emph{\bibinfo{journal}{{Geophysical Journal International}}}
  \textbf{\bibinfo{volume}{207}}, \bibinfo{pages}{29--46}
  (\bibinfo{year}{2016}).

\bibitem{542319:14121529}
\bibinfo{author}{Nakano, M.}, \bibinfo{author}{Sugiyama, D.},
  \bibinfo{author}{Hori, T.}, \bibinfo{author}{Kuwatani, T.} \&
  \bibinfo{author}{Tsuboi, S.}
\newblock \bibinfo{title}{{Discrimination of seismic signals from earthquakes
  and tectonic tremor by applying a convolutional neural network to running
  spectral images}}.
\newblock \emph{\bibinfo{journal}{{Seismological Research Letters}}}
  \textbf{\bibinfo{volume}{90}}, \bibinfo{pages}{530--538}
  (\bibinfo{year}{2019}).

\bibitem{542319:14121530}
\bibinfo{author}{McBrearty, I.~W.}, \bibinfo{author}{Delorey, A.~A.} \&
  \bibinfo{author}{Johnson, P.~A.}
\newblock \bibinfo{title}{{Pairwise association of seismic arrivals with
  convolutional neural networks}}.
\newblock \emph{\bibinfo{journal}{{Seismological Research Letters}}}
  \textbf{\bibinfo{volume}{90}}, \bibinfo{pages}{503--509}
  (\bibinfo{year}{2019}).

\bibitem{542319:14121531}
\bibinfo{author}{McBrearty, I.~W.}, \bibinfo{author}{Gomberg, J.},
  \bibinfo{author}{Delorey, A.~A.} \& \bibinfo{author}{Johnson, P.~A.}
\newblock \bibinfo{title}{{Earthquake Arrival Association with Backprojection
  and Graph TheoryEarthquake Arrival Association with Backprojection and Graph
  Theory}}.
\newblock \emph{\bibinfo{journal}{{Bulletin of the Seismological Society of
  America}}} .

\bibitem{542319:14121532}
\bibinfo{author}{Ross, Z.~E.}, \bibinfo{author}{Yue, Y.},
  \bibinfo{author}{Meier, M.-A.}, \bibinfo{author}{Hauksson, E.} \&
  \bibinfo{author}{Heaton, T.~H.}
\newblock \bibinfo{title}{{PhaseLink: A deep learning approach to seismic phase
  association}}.
\newblock \emph{\bibinfo{journal}{{Journal of Geophysical Research: Solid
  Earth}}} \textbf{\bibinfo{volume}{124}}, \bibinfo{pages}{856--869}
  (\bibinfo{year}{2019}).

\bibitem{542319:14121535}
\bibinfo{author}{Lomax, A.}, \bibinfo{author}{Michelini, A.} \&
  \bibinfo{author}{Jozinovi\'{c}, D.}
\newblock \bibinfo{title}{{An investigation of rapid earthquake
  characterization using single-station waveforms and a convolutional neural
  network}}.
\newblock \emph{\bibinfo{journal}{{Seismological Research Letters}}}
  \textbf{\bibinfo{volume}{90}}, \bibinfo{pages}{517--529}
  (\bibinfo{year}{2019}).

\bibitem{542319:14403038}
\bibinfo{author}{Kriegerowski, M.}, \bibinfo{author}{Petersen, G.~M.},
  \bibinfo{author}{Vasyura-Bathke, H.} \& \bibinfo{author}{Ohrnberger, M.}
\newblock \bibinfo{title}{{A deep convolutional neural network for localization
  of clustered earthquakes based on multistation full waveforms}}.
\newblock \emph{\bibinfo{journal}{{Seismological Research Letters}}}
  \textbf{\bibinfo{volume}{90}}, \bibinfo{pages}{510--516}
  (\bibinfo{year}{2018}).

\bibitem{542319:12461547}
\bibinfo{author}{Ghahramani, Z.}
\newblock \bibinfo{title}{{Probabilistic machine learning and artificial
  intelligence}}.
\newblock \emph{\bibinfo{journal}{{Nature}}} \textbf{\bibinfo{volume}{521}},
  \bibinfo{pages}{452} (\bibinfo{year}{2015}).

\bibitem{542319:12459951}
\bibinfo{author}{Gal, Y.} \& \bibinfo{author}{Ghahramani, Z.}
\newblock \bibinfo{title}{{Dropout as a bayesian approximation: Representing
  model uncertainty in deep learning}}.
\newblock In \emph{\bibinfo{booktitle}{{international conference on machine
  learning}}}, \bibinfo{pages}{1050--1059} (\bibinfo{year}{2016}).

\bibitem{542319:14121746}
\bibinfo{author}{Minson, S.~E.} \emph{et~al.}
\newblock \bibinfo{title}{{Crowdsourced earthquake early warning}}.
\newblock \emph{\bibinfo{journal}{{Science advances}}}
  \textbf{\bibinfo{volume}{1}}, \bibinfo{pages}{e1500036}
  (\bibinfo{year}{2015}).

\bibitem{542319:14404950}
\bibinfo{author}{Mousavi, S.~M.}, \bibinfo{author}{Sheng, Y.},
  \bibinfo{author}{Zhu, W.} \& \bibinfo{author}{Beroza, G.}
\newblock \bibinfo{title}{{STanford EArthquake Dataset (STEAD): A Global Data
  Set of Seismic Signals for AI}}.
\newblock \emph{\bibinfo{journal}{{IEEE Access}}}
  \urlprefix\url{10.1109/ACCESS.2019.2947848}.

\bibitem{542319:14177663}
\bibinfo{author}{Ringler, A.~T.}, \bibinfo{author}{Hutt, C.~R.},
  \bibinfo{author}{Persefield, K.} \& \bibinfo{author}{Gee, L.~S.}
\newblock \bibinfo{title}{{Seismic station installation orientation errors at
  ANSS and IRIS/USGS stations}}.
\newblock \emph{\bibinfo{journal}{{Seismological Research Letters}}}
  \textbf{\bibinfo{volume}{84}}, \bibinfo{pages}{926--931}
  (\bibinfo{year}{2013}).

\bibitem{542319:14177664}
\bibinfo{author}{Ekstr{\"o}m, G.} \& \bibinfo{author}{Busby, R.~W.}
\newblock \bibinfo{title}{{Measurements of seismometer orientation at USArray
  transportable array and backbone stations}}.
\newblock \emph{\bibinfo{journal}{{Seismological Research Letters}}}
  \textbf{\bibinfo{volume}{79}}, \bibinfo{pages}{554--561}
  (\bibinfo{year}{2008}).

\bibitem{542319:14404347}
\bibinfo{author}{Hara, K.}, \bibinfo{author}{Vemulapalli, R.} \&
  \bibinfo{author}{Chellappa, R.}
\newblock \bibinfo{title}{{Designing deep convolutional neural networks for
  continuous object orientation estimation}}.
\newblock \emph{\bibinfo{journal}{{arXiv preprint arXiv:1702.01499}}}
  (\bibinfo{year}{2017}).

\bibitem{542319:14404468}
\bibinfo{author}{Roberts, R.~G.}, \bibinfo{author}{Christoffersson, A.} \&
  \bibinfo{author}{Cassidy, F.}
\newblock \bibinfo{title}{{Real-time event detection, phase identification and
  source location estimation using single station three-component seismic
  data}}.
\newblock \emph{\bibinfo{journal}{{Geophysical Journal International}}}
  \textbf{\bibinfo{volume}{97}}, \bibinfo{pages}{471--480}
  (\bibinfo{year}{1989}).

\bibitem{542319:14404511}
\bibinfo{author}{Cichowicz, A.}
\newblock \bibinfo{title}{{An automatic S-phase picker}}.
\newblock \emph{\bibinfo{journal}{{Bulletin of the Seismological Society of
  America}}} \textbf{\bibinfo{volume}{83}}, \bibinfo{pages}{180--189}
  (\bibinfo{year}{1993}).

\bibitem{542319:14404592}
\bibinfo{author}{Magotra, N.}, \bibinfo{author}{Ahmed, N.} \&
  \bibinfo{author}{Chael, E.}
\newblock \bibinfo{title}{{Single-station seismic event detection and
  location}}.
\newblock \emph{\bibinfo{journal}{{IEEE Transactions on Geoscience and Remote
  Sensing}}} \textbf{\bibinfo{volume}{27}}, \bibinfo{pages}{15--23}
  (\bibinfo{year}{1989}).

\bibitem{542319:14404634}
\bibinfo{author}{B{\"{o}}se, M.} \emph{et~al.}
\newblock \bibinfo{title}{{A probabilistic framework for single-station
  location of seismicity on Earth and Mars}}.
\newblock \emph{\bibinfo{journal}{{Physics of the Earth and Planetary
  Interiors}}} \textbf{\bibinfo{volume}{262}}, \bibinfo{pages}{48--65}
  (\bibinfo{year}{2017}).

\bibitem{542319:14404635}
\bibinfo{author}{Abercrombie, R.~E.}
\newblock \bibinfo{title}{{Earthquake locations using single-station deep
  borehole recordings: Implications for microseismicity on the San Andreas
  fault in southern California}}.
\newblock \emph{\bibinfo{journal}{{Journal of Geophysical Research: Solid
  Earth}}} \textbf{\bibinfo{volume}{100}}, \bibinfo{pages}{24003--24014}
  (\bibinfo{year}{1995}).

\bibitem{542319:12368995}
\bibinfo{author}{Kanamori, H.}
\newblock \bibinfo{title}{{Quantification of earthquakes}}.
\newblock \emph{\bibinfo{journal}{{Nature}}} \textbf{\bibinfo{volume}{271}},
  \bibinfo{pages}{411} (\bibinfo{year}{1978}).

\bibitem{542319:14404645}
\bibinfo{author}{Lockman, A.~B.} \& \bibinfo{author}{Allen, R.~M.}
\newblock \bibinfo{title}{{Single-station earthquake characterization for early
  warning}}.
\newblock \emph{\bibinfo{journal}{{Bulletin of the Seismological Society of
  America}}} \textbf{\bibinfo{volume}{95}}, \bibinfo{pages}{2029--2039}
  (\bibinfo{year}{2005}).

\bibitem{542319:12461322}
\bibinfo{author}{MacKay, D.~J.}
\newblock \bibinfo{title}{{A practical Bayesian framework for backpropagation
  networks}}.
\newblock \emph{\bibinfo{journal}{{Neural computation}}}
  \textbf{\bibinfo{volume}{4}}, \bibinfo{pages}{448--472}
  (\bibinfo{year}{1992}).

\bibitem{542319:12463129}
\bibinfo{author}{Graves, A.}
\newblock \bibinfo{title}{{Practical variational inference for neural
  networks}}.
\newblock In \emph{\bibinfo{booktitle}{{Advances in neural information
  processing systems}}}, \bibinfo{pages}{2348--2356} (\bibinfo{year}{2011}).

\bibitem{542319:12463130}
\bibinfo{author}{Blundell, C.}, \bibinfo{author}{Cornebise, J.},
  \bibinfo{author}{Kavukcuoglu, K.} \& \bibinfo{author}{Wierstra, D.}
\newblock \bibinfo{title}{{Weight uncertainty in neural networks}}.
\newblock \emph{\bibinfo{journal}{{arXiv preprint arXiv:1505.05424}}}
  (\bibinfo{year}{2015}).

\bibitem{542319:12463131}
\bibinfo{author}{Gal, Y.} \& \bibinfo{author}{Ghahramani, Z.}
\newblock \bibinfo{title}{{Bayesian convolutional neural networks with
  Bernoulli approximate variational inference}}.
\newblock \emph{\bibinfo{journal}{{arXiv preprint arXiv:1506.02158}}}
  (\bibinfo{year}{2015}).

\bibitem{542319:12370981}
\bibinfo{author}{Srivastava, N.}, \bibinfo{author}{Hinton, G.},
  \bibinfo{author}{Krizhevsky, A.}, \bibinfo{author}{Sutskever, I.} \&
  \bibinfo{author}{Salakhutdinov, R.}
\newblock \bibinfo{title}{{Dropout: a simple way to prevent neural networks
  from overfitting}}.
\newblock \emph{\bibinfo{journal}{{The journal of machine learning research}}}
  \textbf{\bibinfo{volume}{15}}, \bibinfo{pages}{1929--1958}
  (\bibinfo{year}{2014}).

\bibitem{542319:12464545}
\bibinfo{author}{Kendall, A.}, \bibinfo{author}{Badrinarayanan, V.} \&
  \bibinfo{author}{Cipolla, R.}
\newblock \bibinfo{title}{{Bayesian segnet: Model uncertainty in deep
  convolutional encoder-decoder architectures for scene understanding}}.
\newblock \emph{\bibinfo{journal}{{arXiv preprint arXiv:1511.02680}}}
  (\bibinfo{year}{2015}).

\bibitem{542319:12480069}
\bibinfo{author}{Damianou, A.} \& \bibinfo{author}{Lawrence, N.}
\newblock \bibinfo{title}{{Deep gaussian processes}}.
\newblock In \emph{\bibinfo{booktitle}{{Artificial Intelligence and
  Statistics}}}, \bibinfo{pages}{207--215} (\bibinfo{year}{2013}).

\bibitem{542319:12459953}
\bibinfo{author}{Kendall, A.} \& \bibinfo{author}{Gal, Y.}
\newblock \bibinfo{title}{{What uncertainties do we need in bayesian deep
  learning for computer vision?}}
\newblock In \emph{\bibinfo{booktitle}{{Advances in neural information
  processing systems}}}, \bibinfo{pages}{5574--5584} (\bibinfo{year}{2017}).

\bibitem{542319:14121919}
\bibinfo{author}{Hochreiter, S.} \& \bibinfo{author}{Schmidhuber, J.}
\newblock \bibinfo{title}{{Long short-term memory}}.
\newblock \emph{\bibinfo{journal}{{Neural computation}}}
  \textbf{\bibinfo{volume}{9}}, \bibinfo{pages}{1735--1780}
  (\bibinfo{year}{1997}).

\bibitem{542319:14121961}
\bibinfo{author}{Chung, J.}, \bibinfo{author}{Gulcehre, C.},
  \bibinfo{author}{Cho, K.} \& \bibinfo{author}{Bengio, Y.}
\newblock \bibinfo{title}{{Empirical evaluation of gated recurrent neural
  networks on sequence modeling}}.
\newblock \emph{\bibinfo{journal}{{arXiv preprint arXiv:1412.3555}}}
  (\bibinfo{year}{2014}).

\bibitem{542319:12369449}
\bibinfo{author}{Bai, S.}, \bibinfo{author}{Kolter, J.~Z.} \&
  \bibinfo{author}{Koltun, V.}
\newblock \bibinfo{title}{{An empirical evaluation of generic convolutional and
  recurrent networks for sequence modeling}}.
\newblock \emph{\bibinfo{journal}{{arXiv preprint arXiv:1803.01271}}}
  (\bibinfo{year}{2018}).

\bibitem{542319:12370832}
\bibinfo{author}{van~den Oord, A.} \emph{et~al.}
\newblock \bibinfo{title}{{Wavenet: A generative model for raw audio}}.
\newblock \emph{\bibinfo{journal}{{arXiv preprint arXiv:1609.03499}}}
  (\bibinfo{year}{2016}).

\bibitem{542319:12370874}
\bibinfo{author}{Yu, F.} \& \bibinfo{author}{Koltun, V.}
\newblock \bibinfo{title}{{Multi-scale context aggregation by dilated
  convolutions}}.
\newblock \emph{\bibinfo{journal}{{arXiv preprint arXiv:1511.07122}}}
  (\bibinfo{year}{2015}).

\bibitem{542319:12370984}
\bibinfo{author}{He, K.}, \bibinfo{author}{Zhang, X.}, \bibinfo{author}{Ren,
  S.} \& \bibinfo{author}{Sun, J.}
\newblock \bibinfo{title}{{Deep residual learning for image recognition}}.
\newblock In \emph{\bibinfo{booktitle}{{Proceedings of the IEEE conference on
  computer vision and pattern recognition}}}, \bibinfo{pages}{770--778}
  (\bibinfo{year}{2016}).

\bibitem{542319:12370939}
\bibinfo{author}{Nair, V.} \& \bibinfo{author}{Hinton, G.~E.}
\newblock \bibinfo{title}{{Rectified linear units improve restricted boltzmann
  machines}}.
\newblock In \emph{\bibinfo{booktitle}{{Proceedings of the 27th international
  conference on machine learning (ICML-10)}}}, \bibinfo{pages}{807--814}
  (\bibinfo{year}{2010}).

\end{thebibliography}

\section*{Authors' contributions}S.M.M. designed the project, networks, implemented the software, performed the training and tests, and wrote the manuscript. G.C.B. lead the project and helped with the manuscript. All authors contributed ideas to the project.

\section*{Competing interests}The authors declare any competing interests. 

\newpage 

\end{document}